\documentclass[article,english,11pt]{article}

\usepackage{amscd,amsmath,amssymb,amsfonts,latexsym,mathrsfs,amsthm}
\usepackage{graphicx} 
\usepackage{setspace} 
\usepackage{graphics,epsfig}
\usepackage{sectsty}
\usepackage{natbib}
\usepackage[english]{babel}
\usepackage{amsmath}
\usepackage{amsfonts}
\usepackage{amssymb,amsthm}
\usepackage{bm}
\usepackage{mathrsfs}
\usepackage{subfigure}
\usepackage[usenames]{color}
\usepackage{rotating}
\makeatother

\begin{document}

\title{A Bayesian Beta Markov Random Field Calibration of the Term Structure of Implied Risk Neutral Densities}

\author{\hspace{-18pt}Roberto Casarin\setcounter{footnote}{1}\footnotemark{} \hspace{50pt} Fabrizio Leisen\setcounter{footnote}{2}\footnotemark{}\\
German Molina\setcounter{footnote}{3}\footnotemark{}\hspace{50pt} Enrique ter Horst\setcounter{footnote}{4}\footnotemark{}\hspace{3pt}\setcounter{footnote}{5}
\footnote{Corresponding author: fabrizio.leisen@gmail.com. Author names are in alphabetical order.}\\
\\
{\centering \small\setcounter{footnote}{1}\footnotemark{}\hspace{1pt} University Ca' Foscari, Italy}\\
{\centering \small\setcounter{footnote}{2}\footnotemark{}\hspace{1pt} University of Kent, U.K.}\\
{\centering \small\setcounter{footnote}{3}\footnotemark{}\hspace{1pt} Idalion Capital Group, U.S.}\\
{\centering \small\setcounter{footnote}{4}\footnotemark{}\hspace{1pt} CESA \& IESA, Colombia \& Venezuela}
}

\date{}

\newtheorem{theo}{Theorem}
\newtheorem{defi}{Definition}
\newtheorem{coro}{Corollary}
\newtheorem{prop}{Proposition}[section]
\theoremstyle{remark}
\newtheorem*{rem}{{\bf Remark}}
\newtheorem{lemma}{Lemma}
\newtheorem{proper}{Property}[section]
\newtheorem{algo}{Algorithm}

\maketitle

\begin{abstract}
We build on the work in \cite{FacKin90}, and propose a more general calibration model for implied risk neutral densities. Our model allows for the joint calibration of a set of densities at different maturities and dates through a Bayesian dynamic Beta Markov Random Field. Our approach allows for possible time dependence between densities with the same maturity, and for dependence across maturities at the same point in time. This approach to the problem encompasses model flexibility, parameter parsimony and, more importantly, information pooling across densities.
\end{abstract}

\section{Introduction}
In financial mathematics, it is common to model stock prices as a Geometric Brownian Motion (GBM), with mean drift equal to $\mu$
under the physical probability measure $\mathbb{P}$. In order to price options on the underlying asset, one has to
perform a change of measure to the asset process in order to make it risk neutral, meaning that it makes all investors neutral with respect to risk preferences. Such a probability measure is denoted as $\mathbb{Q}$ \citep{DelbaenSchachermayer11}. In general parametric
stochastic process models, the mathematical problem of performing a change of measure from $\mathbb{P}$ to $\mathbb{Q}$ poses technical problems mainly due to the non-existence of $\mathbb{Q}$ or its non-uniqueness
\citep{DelbaenSchachermayer11,BoyarchenkoLevendorskii02}. When
departing from the regular GBM to the jump diffusion or geometric L\'{e}vy processes setup \citep{TankovCont03}, uniqueness of $\mathbb{Q}$ is not guaranteed, and
several methods such as the Esscher transform are used to circumvent these limitations \citep{Esscher32,GerberShiu94}.

The economic literature has shown an increasing interest in
nonparametric implied risk neutral densities (\cite{FacKin90,WanNiLai11}), since they both allow gauging what the economic agents think about the future, their economic expectations \citep{BlissPanigirtzoglou04,RodriguezterHorst08}, and also provide superior estimates of such risk-neutral densities \citep{WanNiLai11}.  The \cite{FacKin90}
calibration procedure of risk neutral densities, extracted from
derivative prices on the basis of observations on a variable of
interest such as the underlying, allows us to obtain a density
forecast for such a variable. Density forecast is now widely used in many applied economic contexts, and, more in particular, nonparametric calibration of implied risk neutral densities is now used in macroeconomics to generate predictions on inflation and interest rates (see \cite{BhaChi00}, \cite{CarCraMel05}, \cite{devnos12},
\cite{VergotePuigvert10}, \cite{VeselaPuigvert13}, \cite{sihvah14}).

Our contribution provides a dynamic estimation of the
Radon-Nikodym derivative (\cite{Radon}) that allows us to move from a nonparametric estimation of $\mathbb{Q}$ to a nonparametric estimation of $\mathbb{P}$. This last result provides a natural modelling framework for the term structure of the implied nonparametric risk neutral and physical probability distributions, which accounts for the possible dependence between the \emph{Probability Integral Transforms}\footnote{A Probability Integral Transform is defined by a given realization of a random variable $x_{t}$ and as $\text{PIT}_{t}=\int_{-\infty}^{x_{t}}f(y)dy$.}  (PIT) at different
maturities and different dates for a given maturity, whereas previous calibration approaches lack this generality, as they generally do not take advantage of the different sources of dependence between information sources. 

Since the PITs belong to the unit interval, our calibration approach makes use of beta densities as suggested by \cite{FacKin90}. However, we extend their approach in our paper to the multi-maturity, multi-period setting. In order to account for time and cross-maturity dependence, we propose a random field model with beta densities, which fits well-known features of the data regarding dependencies. We make some general assumptions on the time (lags) and spatial (neighbour system) structure that are needed to obtain a parsimonious model. 
We provide a proper Bayesian inference framework, that allow us to include parameter uncertainty in the density calibration. Moreover, the use of hierarchical prior distributions allows us not only to avoid potential over-fitting due to over-parameterization, but also to achieve different degrees of information pooling across maturities.

The use of densities for predicting quantities of interest is now
common in economics and finance, and many recent papers focus on the
combination and the calibration of the predictive densities. Optimal
linear pool of densities is considered in \cite{HallMitchell2007},
\cite{GewekeAmisano2011}, while more general approaches to density
combination are considered in \cite{BilCasRavVan13}, \cite{MitKap13}
and \cite{GneRan13}. Modelling the time evolution of the optimal
combination of predictive densities is one of the challenging issues
tackled in these papers. The issue of calibrating densities is
considered instead in \cite{GneRafWesGol05} and \cite{GneRan13}, which
also propose the use of beta densities to achieve a continuous
deformation of the predictive density, and to obtain well calibrated
PITs. A well calibrated PIT is defined as one where the calibration
function allows you to obtain a cumulative probability distribution of
the observed underlying asset, under the correct target distribution
(physical measure), with a resulting uniform histogram \citep{FacKin90}.
Despite of the presence in the forecasting and financial literature of similar issues, such as the density calibration and combination, the implied risk neutral calibration literature differs substantially from the forecast calibration literature, in that the first one assumes the calibration model is generating the change of measure needed to obtain the physical measure from the risk neutral. Our paper contributes to this stream of literature through a much-needed extension for capturing key features, since it provides a general approach to the joint calibration of densities, allowing for the pooling of information across different predictive densities (the risk neutral densities at different maturities). 

Finally, as an aside note, this paper also
contributes more generally to the literature on modelling data on bounded
domains. Our Bayesian Beta Markov Random Field model approach, and the inference procedure, are original extensions to the multivariate context of the Bayesian Beta models and inferences recently proposed in the statistic literature. See \cite{BraJohThu07} for Bayesian Beta regression and \cite{CasDalLei12} for model selection in Bayesian Beta autoregressive models and the references therein. While we build our proposed approach through a financial application, the model and approach is general and can be used in other areas where more generic, multidimensional calibration approaches are needed.

The paper is organized as follows: Section \ref{Section1} introduces the density calibration problem, and our Bayesian Beta Markov Random Field model for the joint calibration. In Section \ref{Section2}, we discusses the inference difficulties with the proposed model, and develop a numerical procedure for posterior computation. In Section \ref{Section3}, we study the efficiency of our estimation procedure through simulation. In Section \ref{Section4} we provide an application to the Eurodollar currency, while Section \ref{Section5} concludes and discusses potential extensions.

\section{A dynamic calibration model}\label{Section1}
Let $x_{t,\tau_{i}}$, $i=1,\ldots,M$, and $t=1,\ldots,T$, be a set of
underlying realized forward levels (market-implied estimates of the asset level at maturity), available at time $t$, for the different future maturities $\tau_{1},\ldots,\tau_{M}$. Let $F^{Q}_{t,\tau_i}(x)$ and $F^{P}_{t,\tau_i}(x)$ denote the risk neutral and the physical cumulative density functions (cdf) respectively, and $f^{Q}_{t,\tau_i}(x)$ and $f^{P}_{t,\tau_i}(x)$ their probability density functions (pdf).



We assume the following joint deformation model
\begin{equation}
F^{P}_{t}(x_{t,\tau_1},\ldots,x_{t,\tau_M})=C_{t}(F^{Q}_{t,\tau_1}(x_{t,\tau_{1}}),\ldots,F^{Q}_{t,\tau_M}(x_{t,\tau_{M}}))
\end{equation}
where $C_{t}:[0,1]^{M}\rightarrow[0,1]$, $t=1,\ldots,T$, is a sequence of deformation functions. The model can be restated in terms of densities
\begin{equation}
f^{P}_{t,\tau}(x_{t,\tau_1},\ldots,x_{t,\tau_M})=c_{t}(F^{Q}_{t,\tau_1}(x_{t,\tau_{1}}),\ldots,F^{Q}_{t,\tau_M}(x_{t,\tau_{M}}))\prod_{j=1}^{M}f^{Q}_{t,\tau_j}(x_{t,\tau_{j}})
\end{equation}
where $c_t$ is the mixed partial derivative of $C_t$ with respect to all the arguments. Let $y_{jt}=F^{Q}_{t,\tau_{j}}(x_{t,\tau_{j}})$, $j=1,\ldots,M$. Then, in order to model the dependence of the prediction densities at different dates, our modelling assumption is a Beta dynamic Markov Random Field ($\beta$-MRF). Let $E=[0,1]$ be the phase space and $S=\{1,\ldots,M\}$ the finite set of sites (see \cite{Bre99}, ch. 7) corresponding to the different maturities, then our $\beta$-MRF is defined by the following local specification:
\begin{equation}
c_{t}(y_{1t},\ldots,y_{Mt})=\frac{1}{Z_{t}}\prod_{j=1}^{M}c_{jt}(y_{jt}|y_{N(j)})
\end{equation}
where $y_{N(j)}=\{y_{kt},k\in N(j)\subset S\}$ with $N(j)$ a member of the neighbourhood system $N$, $c_{jt}$ represents the $j$-th components of the joint calibration function $c_t$ and $Z_{t}$ is a normalization function which may depend on the parameter of the calibration model and may be unknown for some $\beta$-MRF neighbourhood system specifications.

Modelling the full dependence structure between densities at the different maturities, and allowing for time-change in this structure, may lead to over-parametrized models and consequently to over-fitting problems. Thus, in this paper we consider parsimonious a $\beta$-MRF model. That is, we assume a time-invariant topology $(S,N)$ and focus on two special neighbourhood systems. The first one is a Markov model
\[
N(j)=\left\{
\begin{array}{ll}
\emptyset&\,\hbox{if}\,j=1\\
\{j-1\}&\,\hbox{if}\,j\neq 1\\
\end{array}
\right.
\]
and the second one is a proximity model
\[
N(j)=\left\{
\begin{array}{ll}
\{2\}&\,\hbox{if}\,j=1\\
\{j-1,j+1\}&\,\hbox{if}\,j\neq 1,M\\
\{M-1\}&\,\hbox{if}\,j=M\\
\end{array}
\right.
\]
connecting each density with the two adjacent densities in terms of maturity. 

Following the standard practice in the financial calibration literature (e.g., see \cite{FacKin90})  we assume that the $j$-th component of the joint calibration function is the probability density function of a beta distribution. In order to account for possible time dependence in the PITs, we let the parameter of the beta calibration function of the density at maturity $\tau_{j}$ to depend on the past values of the PITs for the same maturity. Note that this assumption of dependence only on adjacent densities is well supported in financial applications, where, conditional on the PIT of an adjacent maturity, the PIT is independent of PITs of other maturities (since the times-to-maturity are overlapping), and conditional on the closest PIT on a given maturity, the PIT is independent of other PITs for that maturity (basic Markovian assumption for the underlying processes). We use the re-parametrization of beta pdfs used in Bayesian mixture models (e.g., see \cite{RobRou02} and \cite{BouZioMon06}) and Bayesian beta autoregressive processes (e.g., see \cite{CasDalLei12})
\begin{equation}
c_{jt}(y_{jt}|y_{N(j)})=B_{jt}y_{jt}^{\mu_{jt}\gamma_{jt}-1}(1-y_{jt})^{(1-\mu_{jt})\gamma_{jt}-1}
\end{equation}
with
$$
B_{jt}=\frac{\Gamma(\mu_{jt})}{\Gamma(\mu_{jt}\gamma_{jt})\Gamma((1-\mu_{jt})\gamma_{jt})}
$$
and
\begin{eqnarray}
\mu_{jt}&=&\varphi\left(\alpha_{0j}+\sum_{k=1}^{p}\alpha_{kj}y_{t-k,j}+\sum_{k\in N(j)}\beta_{kj}y_{t,k}\right)\\
\gamma_{jt}&=\gamma_{j}&
\end{eqnarray}
with $\varphi:\mathbb{R}\mapsto[0,1]$ a twice differentiable strictly monotonic link function. We assume a logistic function.

\section{Bayesian inference}\label{Section2}
Let $\mathbf{x}_{t}=(x_{t,\tau_{1}},\ldots,x_{t,\tau_{M}})$ be a set of observations for different maturities, and $\mathbf{x}_{p+1:T}=(\mathbf{x}_{p+1},\ldots,\mathbf{x}_{T})$, then the likelihood of the model writes as
\begin{eqnarray}
&&L(\mathbf{x}_{p+1:T}|\boldsymbol{\theta})=\prod_{t=p+1}^{T}f^{P}_{t,\tau}(x_{t,\tau_1},\ldots,x_{t,\tau_M})\\
&&=\prod_{t=p+1}^{T}\frac{1}{Z_{t}}\prod_{j=1}^{M}
B_{jt}(\mu_{jt}\gamma_{j},(1-\mu_{jt})\gamma_{j})\left(F^{Q}_{t,\tau_j}(x_{t,\tau_{j}})\right)^{\mu_{jt}\gamma_{j}-1}\nonumber\\
&&\quad\quad\left(1-F^{Q}_{t,\tau_j}(x_{t,\tau_{j}})\right)^{(1-\mu_{jt})\gamma_{j}-1}f^{Q}_{t,\tau_j}(x_{t,\tau_{j}})\nonumber
\end{eqnarray}
Note that this is a pseudo-likelihood, since we assume that the $p$ initial values of the $\beta$-MRF are known.

In order to complete the description of our Beta Markov Random Field model we assume the following hierarchical specification of the prior distribution. For a given $j$, with $j=1,\ldots,M$, we assume
\begin{eqnarray}
\alpha_{kj}&\overset{i.i.d.}{\sim}&\mathcal{N}(\alpha_{j},s^2_{j})\, k=1,\ldots,p\\
\beta_{kj}&\overset{i.i.d.}{\sim}&\mathcal{N}(\beta_{j},g^2_{j}),\,k=1,\ldots,m_{j},
\end{eqnarray}
For the second level of the hierarchy we assume
\begin{eqnarray}
\gamma_{j}&\overset{i.i.d.}{\sim}&\mathcal{G}a(\xi_1,\xi_2),\, j=1,\ldots,M\\
\alpha_{j}&\overset{i.i.d.}{\sim}&\mathcal{N}(\alpha,s^{2}),\,j=1,\ldots,M\\
\beta_{j}&\overset{i.i.d.}{\sim}&\mathcal{N}(\beta,g^{2}),\,j=1,\ldots,M\\
\alpha&\sim&\mathcal{N}(a,s^{2}_{0})\\
\beta&\sim&\mathcal{N}(b,g^{2}_{0})
\end{eqnarray}
where $m_{j}=\hbox{Card}(N(j))$ is the number of elements of $N(j)$, $\mathcal{G}a(\xi_{1},\xi_{2})$ denotes the gamma distribution with density
$$
f(\gamma|\xi_1,\xi_2)=\frac{1}{\Gamma(\xi_{1})}\gamma^{\xi_1-1}\exp\{-\xi_2\gamma\}\xi_{2}^{\xi_1}
$$
Moreover, in order to design an efficient algorithm for posterior simulation, we re-parametrize $\sigma_{j}=\log(\gamma_{j})$, $j=1,\ldots,M$. We will define the parameter vector $\boldsymbol{\theta}=(\boldsymbol{\theta}_{1},\ldots,\boldsymbol{\theta}_{M},\alpha,\beta)$ where $\boldsymbol{\theta}_{j}=(\boldsymbol{\alpha}_{j},\boldsymbol{\beta}_{j},\sigma_{j},\alpha_j,\beta_j)$, $\boldsymbol{\alpha}_{j}=(\alpha_{0j},\alpha_{1j},\ldots,\alpha_{pj})$ and $\boldsymbol{\beta}_{j}=(\beta_{1j},\ldots,\beta_{m_j j})$. Then the joint probability density function of the prior distribution is
\begin{eqnarray}
&&\!\!\!\!\!f(\boldsymbol{\theta})\propto \exp\Bigg\{
-\frac{1}{2s_{0}^{2}}(\alpha-a)^{2}-\frac{1}{2g_{0}^{2}}(\beta-b)^{2}-\sum_{j=1}^{M}\left(\frac{1}{2s^{2}}(\alpha_{j}-\alpha)^{2}
\right.\\
&&%
+\frac{1}{2g^{2}}(\beta_{j}-\beta)^{2}+\frac{1}{2}(\boldsymbol{\alpha}_{j}-\boldsymbol{\mu}_{j})'S_{j}^{-1}(\boldsymbol{\alpha}_{j}-\boldsymbol{\mu}_{j})\nonumber\\
&&\left.+\frac{1}{2}(\boldsymbol{\beta}_{j}-\boldsymbol{\nu}_{j})'G_{j}^{-1}(\boldsymbol{\beta}_{j}-\boldsymbol{\nu}_{j})\right)\Bigg\}
\prod_{j=1}^{M}\exp\{-\xi_{1}/2\exp(\sigma_{j})\}\exp(\xi_{2}/2\sigma_{j})
\nonumber
\end{eqnarray}
where $\boldsymbol{\mu}_{j}=\alpha_{j}\boldsymbol{\iota}_{(p+1)}$, $\boldsymbol{\nu}_{j}=\beta_{j}\boldsymbol{\iota}_{m_j}$, with $\boldsymbol{\iota}_{n}$ the $n$-dimensional unit vector. The prior covariance matrices are $S_{j}=s^{2}_{j} I_{(p+1)}$ and $G_{j}=g^{2}_{j} I_{m}$, with $I_{n}$ the $n$-dimensional identity matrix.

The joint posterior distribution can be written as
\begin{eqnarray}
&&\pi(\boldsymbol{\theta}|\mathbf{x}_{p+1:T})\propto\exp\left(
-\sum_{t=p+1}^{T}\log Z_{t}-\sum_{t=p+1}^{T}\sum_{j=1}^{M}\log B_{jt}\right.\\
&&\left.%
\quad\quad+\sum_{t=p+1}^{T}\sum_{j=1}^{M}\left(A_{jt}\mu_{jt}+\log(1-F^{Q}_{t,\tau_j}(x_{t,\tau_{j}}))\right)\exp(\sigma_{j})\right)f(\boldsymbol{\theta})\nonumber
\end{eqnarray}
where 
$$
B_{jt}=B_{jt}(\mu_{jt}\exp(\sigma_{j}),(1-\mu_{jt})\exp(\sigma_{j}))
$$
and 
$$
A_{jt}=\log(F^{Q}_{t,\tau_j}(x_{t,\tau_{j}})/(1-F^{Q}_{t,\tau_j}(x_{t,\tau_{j}})))
$$

A major problem with this model is that the normalizing constants $Z_{t}$, $t=p+1,\ldots,T$, in the likelihood function and in the posterior distribution are unknown and possibly depend on the parameters. Thus, samples from $\pi(\boldsymbol{\theta}|\mathbf{x}_{p+1:T})$ cannot be easily obtained with standard MCMC procedures. For instance, the standard MH algorithm  cannot be directly applied because the acceptance probability involves ratios of unknown normalizing constants. In the last two decades, various approximation methods have been proposed in order to circumvent the problem of intractable normalizing constants. Recently \cite{MolPetReeBer06} proposed an auxiliary variable MCMC algorithm, which is a feasible simulation procedure for many models with intractable normalizing constant. The \cite{MolPetReeBer06}'s single auxiliary variable method has been successfully improved by \cite{MurGhaMac06}. They propose the exchange algorithm, which removes the need to estimate the parameter before sampling begins, and has higher acceptance probability than \cite{MolPetReeBer06} 's algorithm. Unfortunately both the single auxiliary variable and the exchange algorithms require exact sampling of the auxiliary variable from its conditional distribution, which can be computationally expensive for many statistical models. An exact simulation algorithm for our $\beta$-MRF model is not available, thus in this paper we follow and alternative route and apply the double MH algorithm proposed by \cite{Lia10}. The double MH avoids the exact simulation step by applying an internal MH step to generate the auxiliary variable.

Assume we are interested in simulating the auxiliary variable $\mathbf{z}_{p+1:T}$ from the conditional distribution $L(\mathbf{z}_{p+1:T}|\boldsymbol{\theta}')$. If the sample is generated by iterating $n$ times a MH algorithm with transition kernel $K_{\boldsymbol{\theta}'}(\mathbf{z}|\mathbf{x})$, then the $n$-step transition probability is
$$
P^{n}_{\boldsymbol{\theta}'}(\mathbf{z}_{p+1:T}|\mathbf{x}_{p+1:T})=
K_{\boldsymbol{\theta}'}(\mathbf{x}_{p+1:T}^{1}|\mathbf{x}_{p+1:T})\cdots K_{\boldsymbol{\theta}'}(\mathbf{z}_{p+1:T}|\mathbf{x}_{p+1:T}^{n-1})
$$
then the acceptance rate of the \cite{MurGhaMac06}'s exchange algorithm writes as
\begin{equation}\label{EqAcc}
\rho(\boldsymbol{\theta},\boldsymbol{\theta}',\mathbf{z}_{p+1:T}|\mathbf{x}_{p+1:T})=\frac{f(\boldsymbol{\theta})q(\boldsymbol{\theta}|\boldsymbol{\theta}',\mathbf{x}_{p+1:T})}{f(\boldsymbol{\theta}')q(\boldsymbol{\theta}'|\boldsymbol{\theta},\mathbf{x}_{p+1:T})}
\frac{L(\mathbf{z}_{p+1:T}|\boldsymbol{\theta})}{L(\mathbf{x}_{p+1:T}|\boldsymbol{\theta})}
\frac{P^{n}_{\boldsymbol{\theta}'}(\mathbf{z}_{p+1:T}|\mathbf{x}_{p+1:T})}{P^{n}_{\boldsymbol{\theta}'}(\mathbf{x}_{p+1:T}|\mathbf{z}_{p+1:T})}
\end{equation}
If we chose $q(\boldsymbol{\theta}|\boldsymbol{\theta}',\mathbf{x}_{p+1:T})$ as a Metropolis transition kernel then the exchange is a MH step with transition $P^{n}_{\boldsymbol{\theta}'}(\mathbf{z}_{p+1:T}|\mathbf{x}_{p+1:T})$ and target distribution $L(\mathbf{z}_{p+1:T}|\boldsymbol{\theta})$, and the acceptance probability in Eq. \ref{EqAcc} becomes
\begin{equation}\label{EqAcc1}
\rho(\boldsymbol{\theta},\boldsymbol{\theta}',\mathbf{z}_{p+1:T}|\mathbf{x}_{p+1:T})=\frac{L(\mathbf{z}_{p+1:T}|\boldsymbol{\theta})}{L(\mathbf{x}_{p+1:T}|\boldsymbol{\theta})}\frac{L(\mathbf{x}_{p+1:T}|\boldsymbol{\theta}')}{L(\mathbf{z}_{p+1:T}|\boldsymbol{\theta}')}
\end{equation}
Assume the current value of the MH chain is $\boldsymbol{\theta}^{(t)}=\boldsymbol{\theta}$, then the double MH sampler iterates over the following steps
\begin{enumerate}
\item Simulate a new sample $\boldsymbol{\theta}'$ from $\pi(\boldsymbol{\theta})$ using a MH algorithm starting with $\boldsymbol{\theta}$.
\item Generate the auxiliary variable $\mathbf{z}_{p+1:T}\sim P^{n}_{\boldsymbol{\theta}'}(\mathbf{z}_{p+1:T}|\mathbf{x}_{p+1:T})$ and accept it with probability $\min\{1,\rho(\boldsymbol{\theta},\boldsymbol{\theta}',\mathbf{z}_{p+1:T}|\mathbf{x}_{p+1:T})\}$ given in Eq. \ref{EqAcc1}
\item Set $\boldsymbol{\theta}^{(t+1)}=\boldsymbol{\theta}'$ if the auxiliary variable is accepted and $\boldsymbol{\theta}_{t+1}=\boldsymbol{\theta}$ otherwise.
\end{enumerate}
As regards to the first MH step in the double MH, we assume a multivariate random-walk proposal, i.e.
$$
\boldsymbol{\theta}^{*}\sim\mathcal{N}(\boldsymbol{\theta}^{(t)},\Lambda)
$$
where $\Lambda$ a $n$-dimensional positive diagonal matrix, with $n=(p+4)M+m+2$. 

Regarding the second MH step we consider a Gibbs sampler which generates samples iteratively from the full conditional distributions of each site. By using the Markov property of our dynamic random field with respect the chosen neighbourhood system, the full conditional distribution of the $j$-th site, conditionally on the remaining sites is a function of the sites in the neighbourhood of $j$, i.e.
\begin{eqnarray}
&&\!\!\!\!\!\!\!\!\pi(x_{t,\tau_{j}}|x_{t,\tau_{j-1}},x_{t,\tau_{j+1}},\boldsymbol{\theta})\propto\\
&&\!\!\!\!\!\!\!\!B_{jt}\left(F^{Q}_{t,\tau_j}(x_{t,\tau_{j}})\right)^{\mu_{jt}\exp(\sigma_j)-1}\left(1-F^{Q}_{t,\tau_j}(x_{t,\tau_{j}})\right)^{(1-\mu_{jt})\exp(\sigma_j)-1}f^{Q}_{t,\tau_j}(x_{t,\tau_{j}})\nonumber\\
&&\!\!\!\!\!\!\!\!\prod_{k=1}^{p}B_{j,t+k}\left(F^{Q}_{t+k,\tau_j}(x_{t+k,\tau_{j}})\right)^{\mu_{j,t+k}\exp(\sigma_j)-1}\left(1-F^{Q}_{t+k,\tau_j}(x_{t+k,\tau_{j}})\right)^{(1-\mu_{j,t+k})\exp(\sigma_j)-1}f^{Q}_{t+k,\tau_j}(x_{t+k,\tau_{j}})\nonumber\\
&&\!\!\!\!\!\!\!\!\prod_{i\in N(j)}
B_{it}\left(F^{Q}_{t,\tau_i}(x_{t,\tau_{i}})\right)^{\mu_{it}\exp(\sigma_i)-1}\left(1-F^{Q}_{t,\tau_i}(x_{t,\tau_{i}})\right)^{(1-\mu_{it})\exp(\sigma_i)-1}f^{Q}_{t,\tau_i}(x_{t,\tau_{i}})\nonumber
\end{eqnarray}
for $t=p+1,\ldots,T$ and $j=1,\ldots,M$. The full conditionals are not easy to simulate from, thus we apply a MH step with proposal distribution
$$
q(x|x_{t,\tau_{j-1}},x_{t,\tau_{j+1}},\boldsymbol{\theta})\propto \left(F^{Q}_{t,\tau_j}(x)\right)^{\mu_{jt}\gamma_{jt}-1}\left(1-F^{Q}_{t,\tau_j}(x)\right)^{(1-\mu_{jt})\gamma_{jt}-1}f^{Q}_{t,\tau_j}(x)
$$
which can be simulated exactly as follows: $y^* \sim \mathcal{B}e(\mu_{jt}\exp(\sigma_{j}),(1-\mu_{jt})\exp(\sigma_{j}))$ and $x^* =F^{Q,-1}_{t,\tau_j}(y^{*})$. This choice of the proposal distribution leads to a simplification of the logarithmic acceptance probability:
$$
\log\tilde{\rho}=\sum_{i\in N(j)}\left(\log B_{it}-\log B_{it}^{*}+A_{it}(\mu_{it}^{*} -\mu_{it})\exp(\sigma_{i})\right)
$$
with $B_{it}^{*}=B_{it}(\mu_{it}^{*}\exp(\sigma_{i}),(1-\mu_{it}^{*})\exp(\sigma_{i}))$ and 
$$
\mu_{it}^{*}=\varphi\left(\alpha_{0i}+\sum_{k=1}^{p}\alpha_{kj}y_{t-k,i}+\sum_{k\in N(i),k\neq j}\beta_{ki}y_{t,k}+\beta_{ji}y^{*}\right)
$$
which follows by the definition of neighbourhood system, that is if $i\in N(j)$ then $j\in N(i)$.
%

\section{Simulation exercise}\label{Section3}
The extraction of parametric and nonparametric risk-neutral densities has been important not only
for traders in order to use this density to price more exotic
derivatives, but also for central bankers as well and policy
makers \citep{Ait03,RouahVainberg07}. Recently, a great deal of interest has grown in predicting both the nonparametric risk-neutral and its physical counterpart simultaneously, as shown in \cite{VeselaPuigvert13} for the 3-month Euribor interest rate,  using the beta calibration function for fixed expirations of the nonparametric risk-neutral density, instead of constant and rolling maturity expirations such as 3,6,9, and 12 months as in
\cite{VergotePuigvert10}. These constant maturity risk-neutral
densities are interpolated in practice from fixed expiration densities as done in \cite{VergotePuigvert10}. We do not need to follow their approach, since we have rolling, fixed time to maturity (as opposed to fixed maturity), option prices and forwards available. In this sense, our approach is more generalizable for over-the-counter markets, where the market centers around (rolling tenor) fixed time-to-maturity points, rather than fixed expiry dates.

\begin{figure}[p]
\centering
\begin{tabular}{cc}
  $\sigma=0.1$&$\sigma=0.2$\\
  \includegraphics[width=170pt, height=150pt, angle=0]{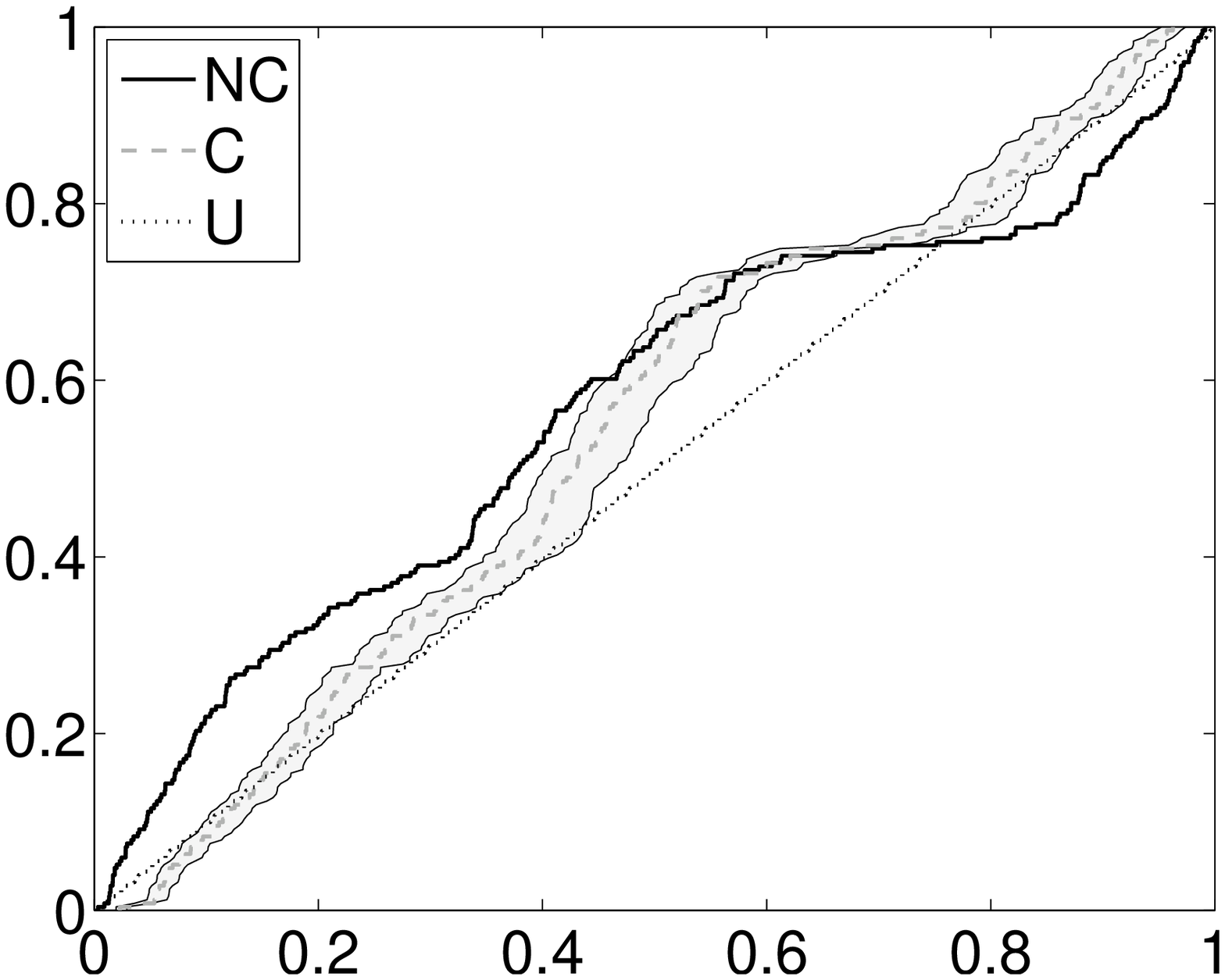}&  \includegraphics[width=170pt, height=150pt, angle=0]{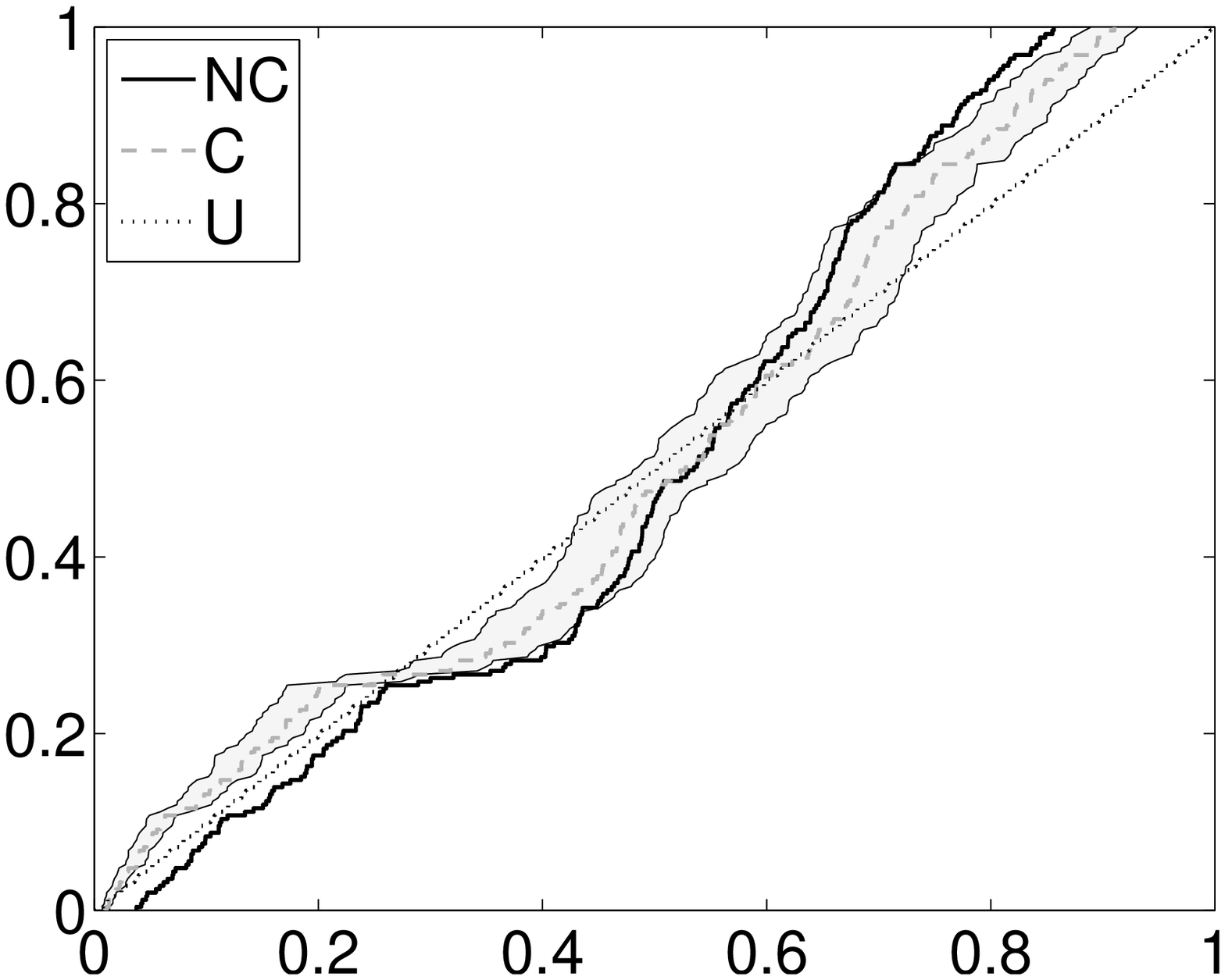}\\
  \includegraphics[width=170pt, height=150pt, angle=0]{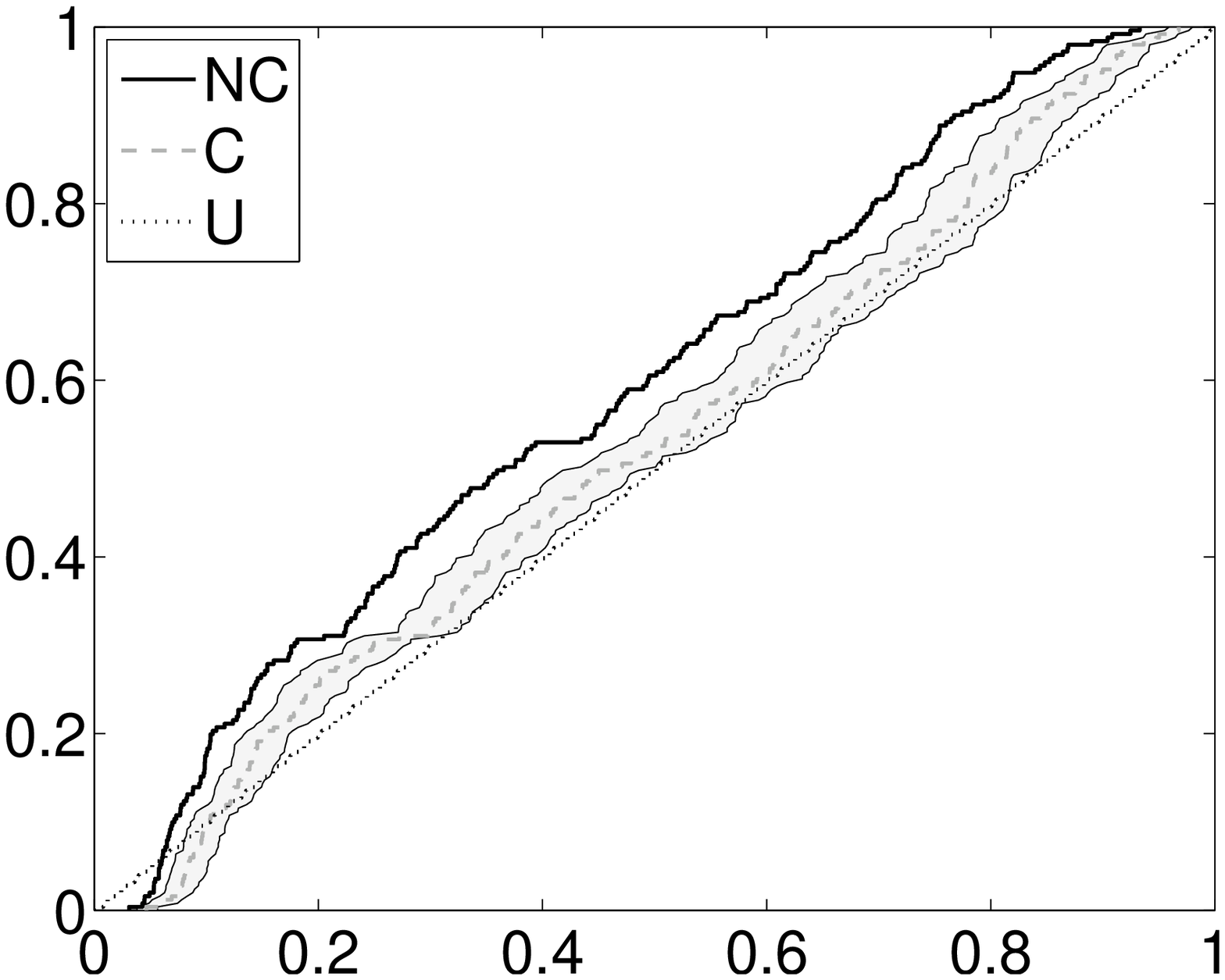}&  \includegraphics[width=170pt, height=150pt, angle=0]{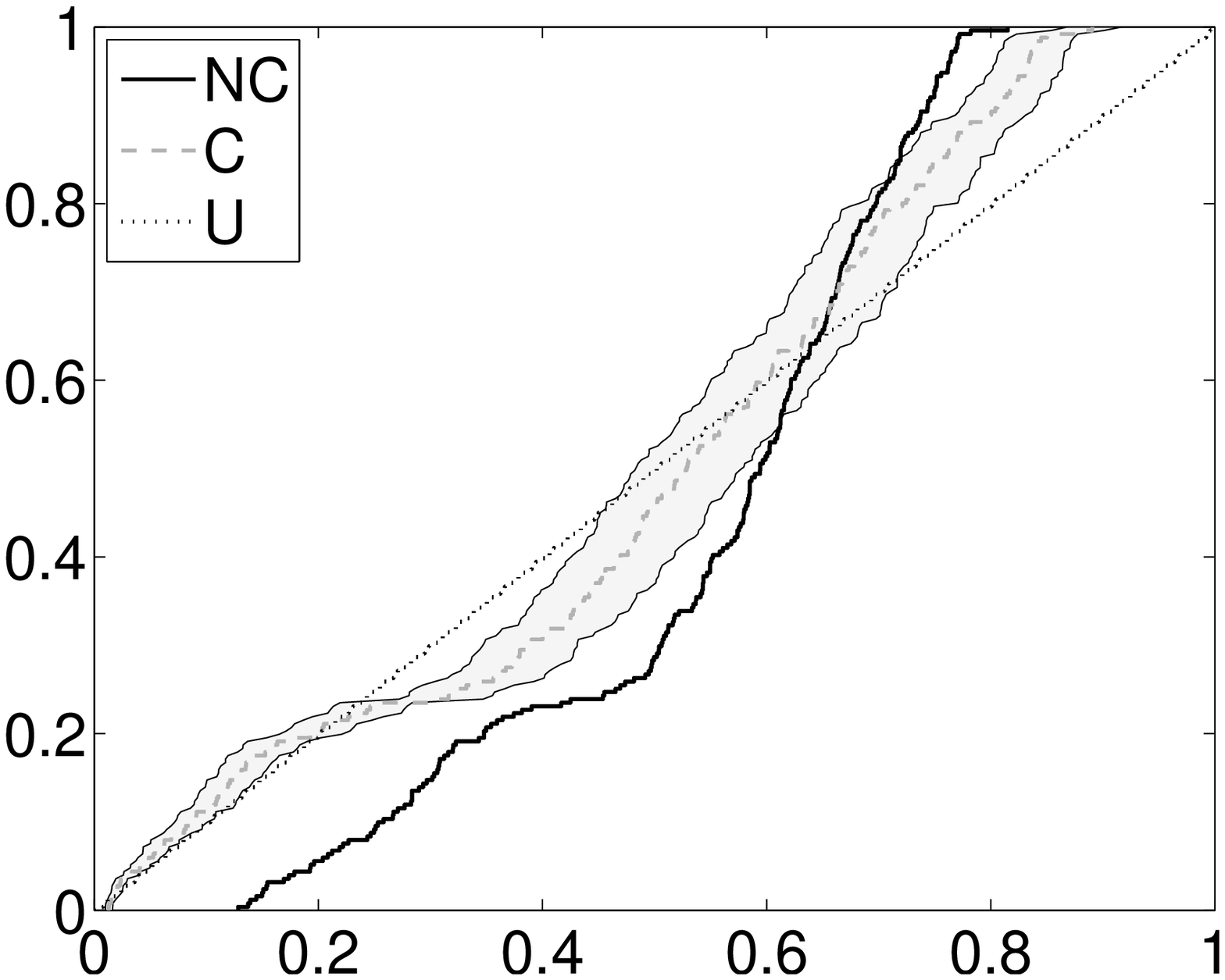}\\
  \includegraphics[width=170pt, height=150pt, angle=0]{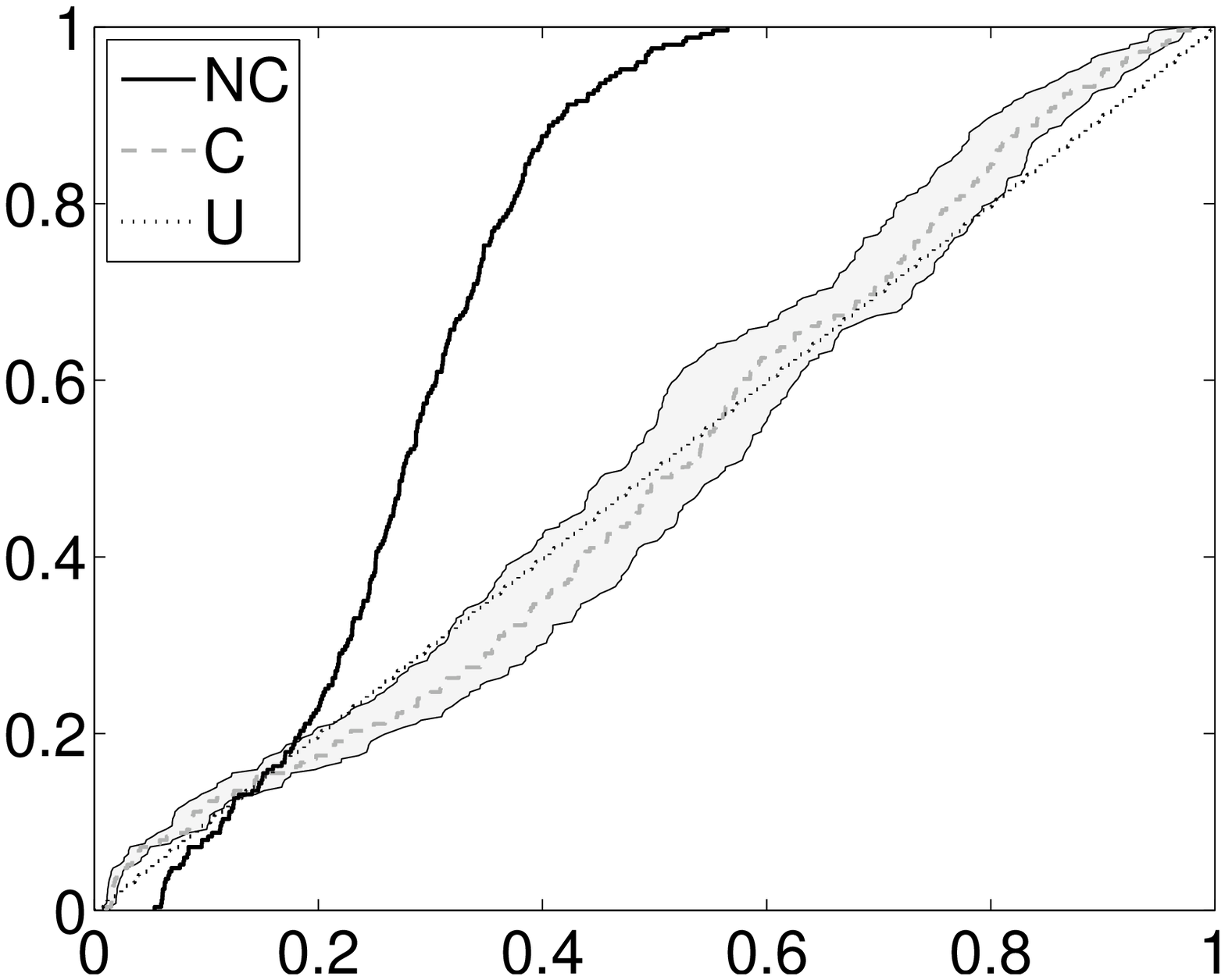}&  \includegraphics[width=170pt, height=150pt, angle=0]{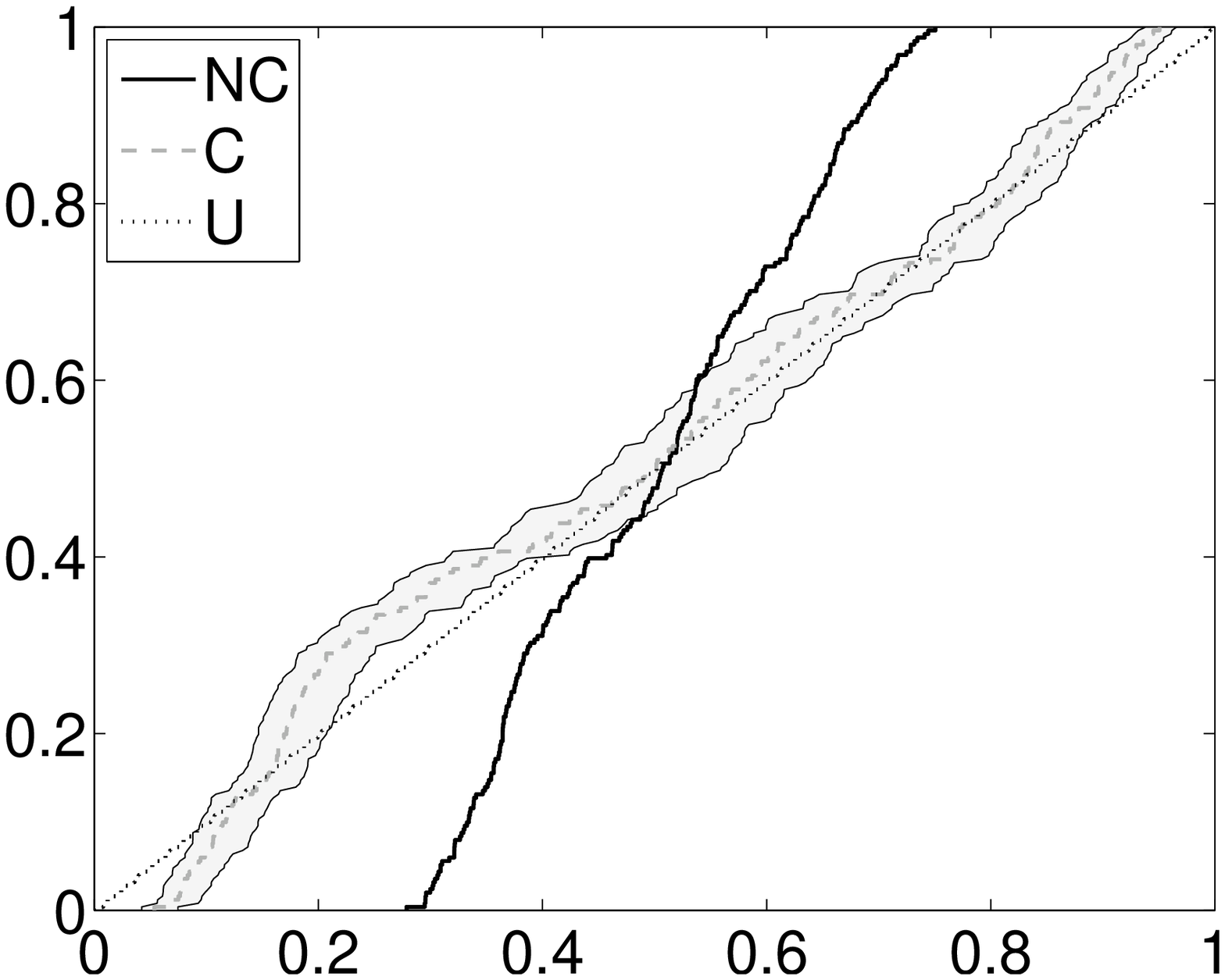}\\
\end{tabular}
  \caption{Non calibrated and calibrated risk neutral distribution for different maturities (rows) and volatility levels (columns).}\label{CalibrationSim}
\end{figure}

In this section we run several simulation exercises to test the
accuracy of our method to produce a calibration function that allows for better assessment of the non-standard features usually encountered in the PIT data, including through mis-estimation of the underlying parameters of the process. This exercise consists of several layers according to
the following sequence:
\begin{itemize}
\item First we produce the simulated data under the physical measure, which will be common to all the simulation exercises. We simulate price paths, with a GBM for the asset price, under the physical measure for 3, 6 and 12 months for a time interval of $T = 2$ years, $\mu=0.20$, $r = 0.05$, $\sigma = 0.15$, $\tau_{1} = 0.25$ (years), $\tau_{2} = 0.5$ (years) and $\tau_{3} = 1$ (year).
\item From that data, we estimate the risk neutral measure, assuming that we incorrectly estimate (grossly) the parameters of this risk neutral measure. For this purpose, we assume two potential scenarios that cover the two extremes:
\begin{enumerate}
\item Overestimation of the volatility of the Brownian Motion: We will assume for the calibration exercise that we overestimate the unknown volatility of the physical process and set $\sigma = 0.20$.
\item Underestimation of the volatility of the Brownian Motion: We will assume for the calibration exercise that we underestimate the unknown volatility of the physical process and set $\sigma = 0.10$.
\item Note that, in both cases, we are using r different than $\mu$.
\end{enumerate}
\item For each of the cases above, and for each of the maturities in the simulation exercise, we compare two curves (Figure
  \eqref{CalibrationSim}):
\begin{enumerate}
\item NC Curve: This is the non-calibrated curve. It simply states the
  shape of the PITs CDF using the risk neutral data, under the stated
  value of the volatility.
\item C Curve: This is the calibrated data using the $\beta$-MRF process using the risk neutral data, under the stated value of the volatility.
\item Note that in both cases we use the same playing field for the volatility, to ensure that the comparison is done solely on the calibration benefits.
\end{enumerate}
\item As a reference, the 45 degree line represents the perfect scenario where the (calibrated) PITs are are uniformly distributed.
\end{itemize}
In order to run this simulation, we assume that the data comes
from a standard process in the financial literature (GBM), with $S_{t}$, $t\in[0,T]$, to model the price of the underlying as in \cite{bs:1973} and \cite{me:1973}, i.e.
\begin{equation}
S_{t}=S_{0}+\int_{0}^{t}S_{u}\mu du+ \int_{0}^{t}S_{u}\sigma dW(u)
\end{equation}
where $W_{t}$, $t\in[0,T]$, is a Wiener process.

We simulate price sample paths under the physical measure for 3, 6  and 12 months for a time interval of $T=2$ years, $\mu=0.20$, $r=0.05$, $\sigma=0.15$, $\tau_{1}=0.25$, $\tau_{2}=0.5$, and $\tau_{3}=1$. 

We also know analytically the risk-neutral densities
of $S_{t+\tau_{j}}$, $j=1,2,3$, conditional on $S_{t}$, which are given by:
\begin{equation}
f^{Q}_{t,\tau_{j}}(S_{t+\tau_{j}}) = \frac{1}{S_{t+\tau_{j}}\sqrt{2\pi\sigma^{2}\tau_{j}}}\exp\left[-\frac{[\log(S_{t+\tau_{j}}/S_{t})-(r-0.5\sigma^{2})\tau_{j}]}{2\sigma^{2}\tau_{j}}\right]\\
\end{equation}
$j=1,2,3$. Once we observe 3 months later a price level of $S_{t+\tau_{1}}$
under the historical measure, then we proceed to compute the 3, 6 and 12 months PITs at time $t$ as follows:
\begin{equation}\label{eq:3mpit}
y_{t,\tau_{j}}=\int_{-\infty}^{S_{t+\tau_{j}}}f^{Q}_{t,\tau_{j}}(S_{t+\tau_{j}})dS_{t+\tau_{j}}
\end{equation}
$j=1,2,3$. The next day at time $t_{1}=t+1$, we recompute the PITs in the same
way as equation \eqref{eq:3mpit}, obtaining a
vector
$\mathbf{x}_{t}=(\mathbf{x}_{t},\mathbf{x}_{t+1},...,\mathbf{x}_{t+T})$
where again
$\mathbf{x}_{s}=(F^{Q}_{s,\tau_{1}}(x_{s,\tau_{1}}),F^{Q}_{s,\tau_{2}}(x_{s,\tau_{2}}),F^{Q}_{s,\tau_{3}}(x_{s,\tau_{3}}))$,
and where the components of $\mathbf{x}_{s}$ will be very likely
correlated, given the overlapping times to maturity. In our simulation exercise we assume that a year has 252
trading days (prices) and that 3 (6 and 12) months correspond to 63 (126 and 252) trading days respectively. 

A uniform marginal distribution of the PITs, assuming that they are not autocorrelated, indicates that there is no need for a calibration function.
A uniform marginal distribution of the PITs, assuming that they are autocorrelated, does not necessarily say anything about the need for a calibration function. There could be cases where the PITs are extremely autocorrelated, and yet display a perfect uniform histogram leading to the wrong conclusion that both the risk neutral and physical measures are both identical.

The source of autocorrelation of the PITs comes from the rolling
nature of the data. Each period t, we obtain a new PIT for each maturity, which is the outcome of the physical process under that given maturity. Since, for a
given maturity $\tau$, we will be producing $\tau\times 252$
overlapping periods (with different levels of overlap), these periods
will share common contributions to each of those PITs. For example, a
3 month PIT with reference point today, and maturity in 65 business
days (3 months), will share 64 business days in common with another PIT,
with reference point tomorrow, and maturity 65 days from tomorrow. This
generates an artificial autocorrelation in the PITs, which is embedded
in any overlapping data. Classical approaches include a mere thinning
of the data (which we do in our simulation exercises) to take only
non-overlapping periods. However, this approach is especially
penalizing on the longer maturities. For example, for maturities of a year, traditional approaches will only collect one data point per year. Our approach is more general, since it takes into account in the modelling the different sources of correlation (including this autocorrelation) between the PITs
through the $\beta$-MRF approach. For two given PITs
$(A,B)$, for which the data driving them is represented by the
combination of the starting points $t_A$, $t_B$, and the maturities
$\tau_A$, $\tau_B$, the overlapping amounts of information contained
in the physical process is the intersection of $[t_A,t_A+\tau_A]\cap [t_B,t_B+\tau_B]$
This information is processed naturally through the $\beta$-MRF approach, which takes into account the two causes of autocorrelation (over time and over neighbors).

We apply our Bayesian $\beta$-MRF calibration model with the following hyper parameter settings $\alpha=0$, $\beta=0$, $s^2_{j}=10$, $g^2_{j}=10$, $s^{2}=100$, and $g=100$. We apply the proposed MCMC algorithm in order to approximate the posterior quantities of interest. In the MCMC algorithm we consider 5,000 iterations after convergence (that is detected after about 2,000 burn-in iterations by applying the \cite{GJ92} convergence diagnostic test statistics). The scale $\Lambda$ of the proposal distribution of the MH step for generating $\boldsymbol{\theta}$ from $q$ was chosen to achieve average acceptance rates between 0.5 and 0.7 for the two MH algorithms (steps 1 and 2), which is a good sign of efficiency for most MCMC algorithms, as suggested, for example, by \cite{Ros11}. This choice can be done 'on-line' through the runs of the algorithm in the burn-in phase.

With regards to Table \eqref{TabMCMC}:
\begin{itemize}
\item $\boldsymbol{\alpha}_{j}$ are the autoregressive parts of the $\beta$-MRF (time factor) – representing the time-dependence.
\item $\boldsymbol{\beta}_{j}$ are the parameters linking the different maturities (maturity factor), representing the cross-maturity dependence.
\end{itemize}
The autocorrelation over time decreases as the
maturity increases. This can be seen in the value of the
corresponding parameters $\alpha$. Additionally, $\beta_1$ and $\beta_2$ represent the correlation parameters of neighboring maturities, before and after respectively. So
$\beta_{12}$ represents the correlation parameters between maturity 1
and maturity 2, while $\beta_{23}$ represents the correlation
parameter between maturity 2 and maturity 3.
Furthermore, panels c and d pool across maturities. This pooling produces an interesting practical approach, because it assumes the same
autoregressive structure over time for the PITs across their
maturities. 

\begin{figure}[p]
\centering
\begin{tabular}{cc}
  $\sigma=0.1$&$\sigma=0.2$\\
  \includegraphics[width=170pt, height=150pt, angle=0]{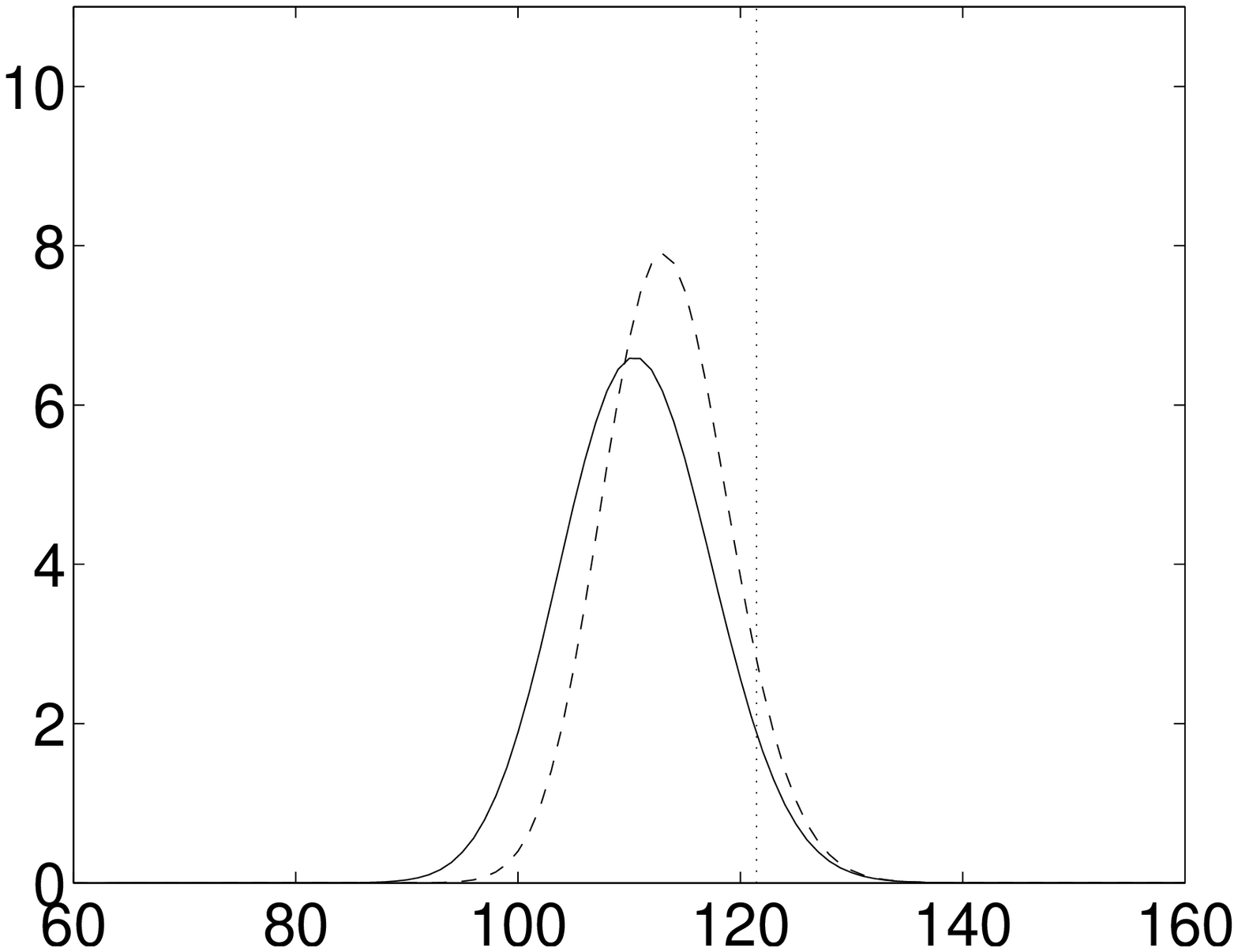}&  \includegraphics[width=170pt, height=150pt, angle=0]{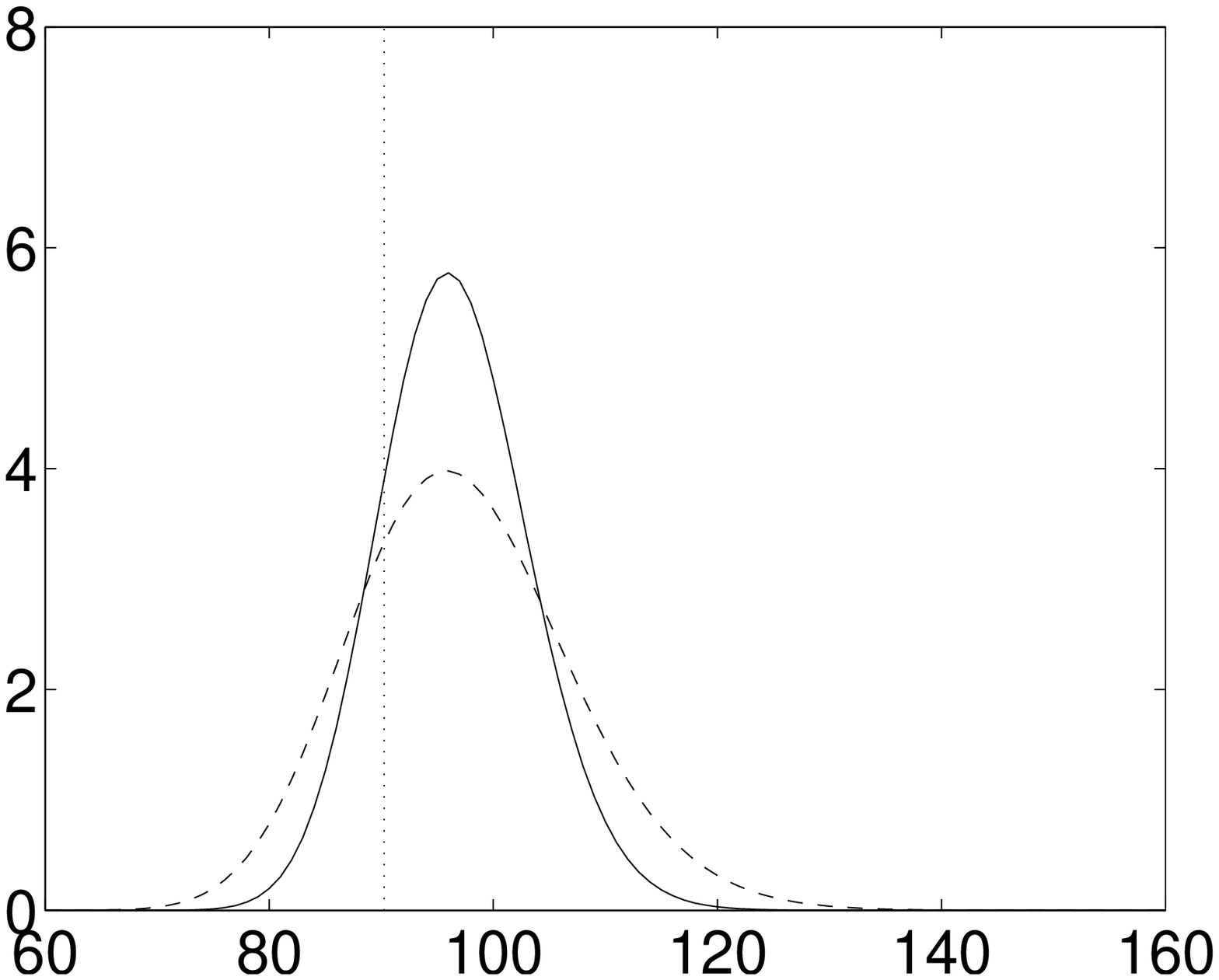}\\
  \includegraphics[width=170pt, height=150pt, angle=0]{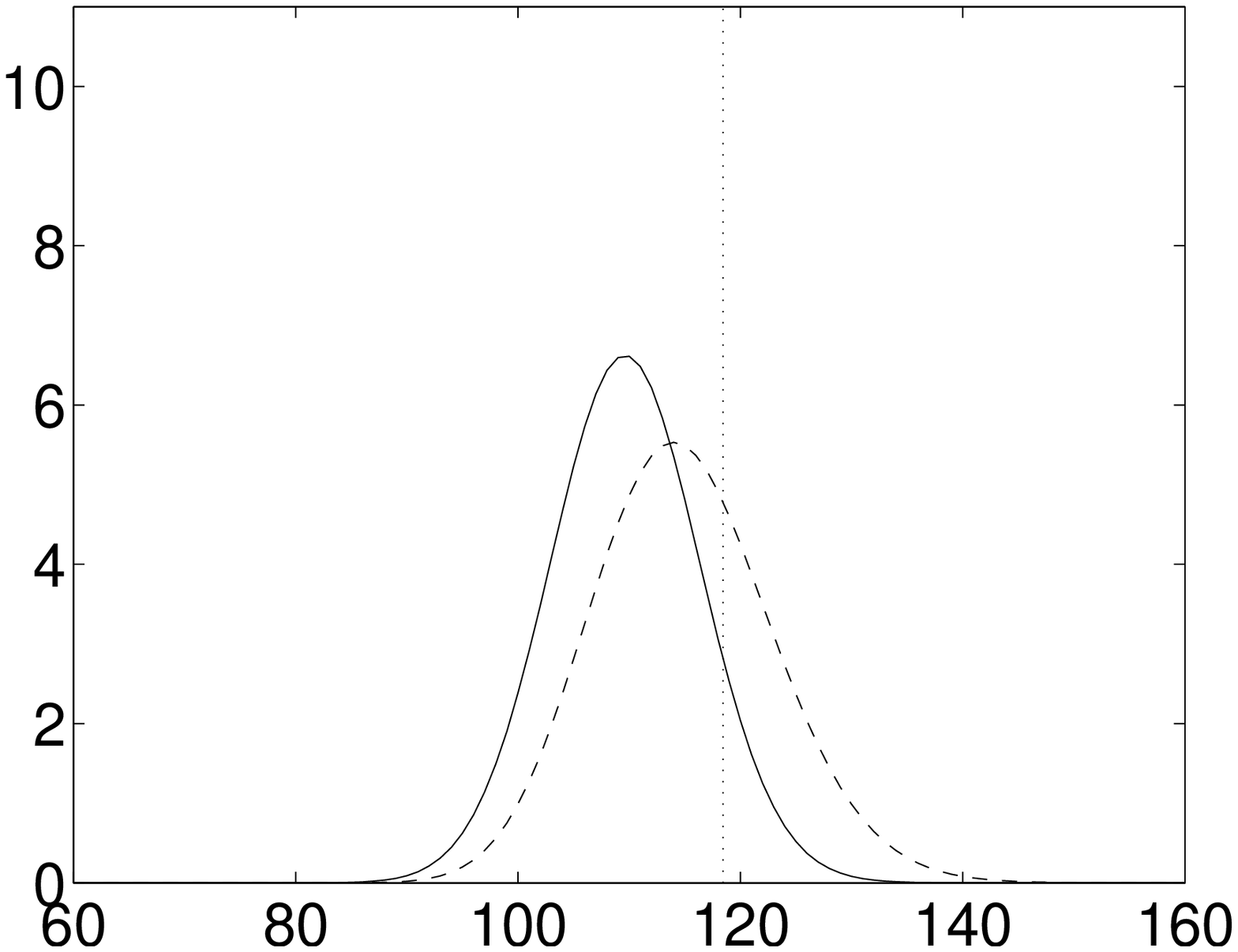}&  \includegraphics[width=170pt, height=150pt, angle=0]{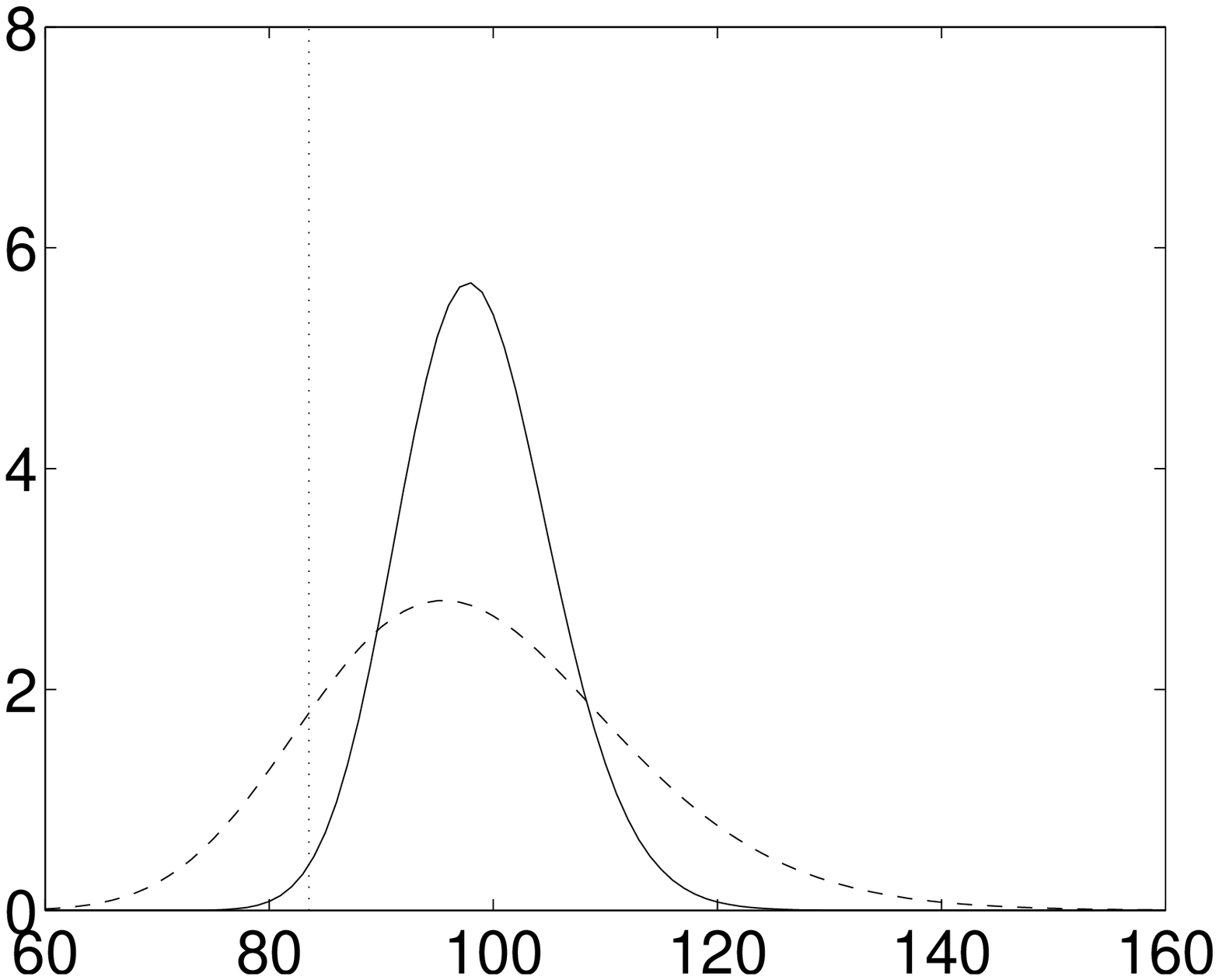}\\
  \includegraphics[width=170pt, height=150pt, angle=0]{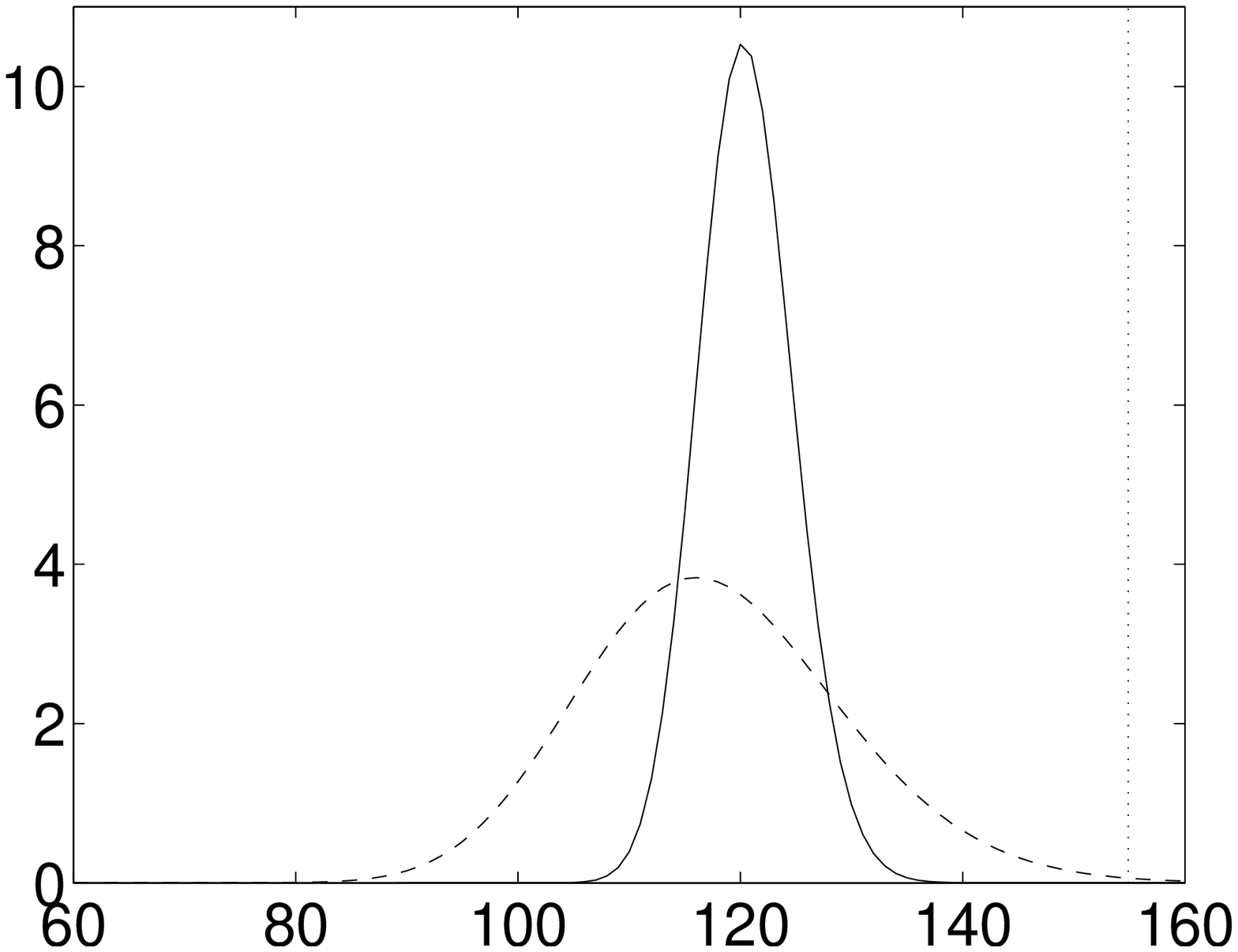}&  \includegraphics[width=170pt, height=150pt, angle=0]{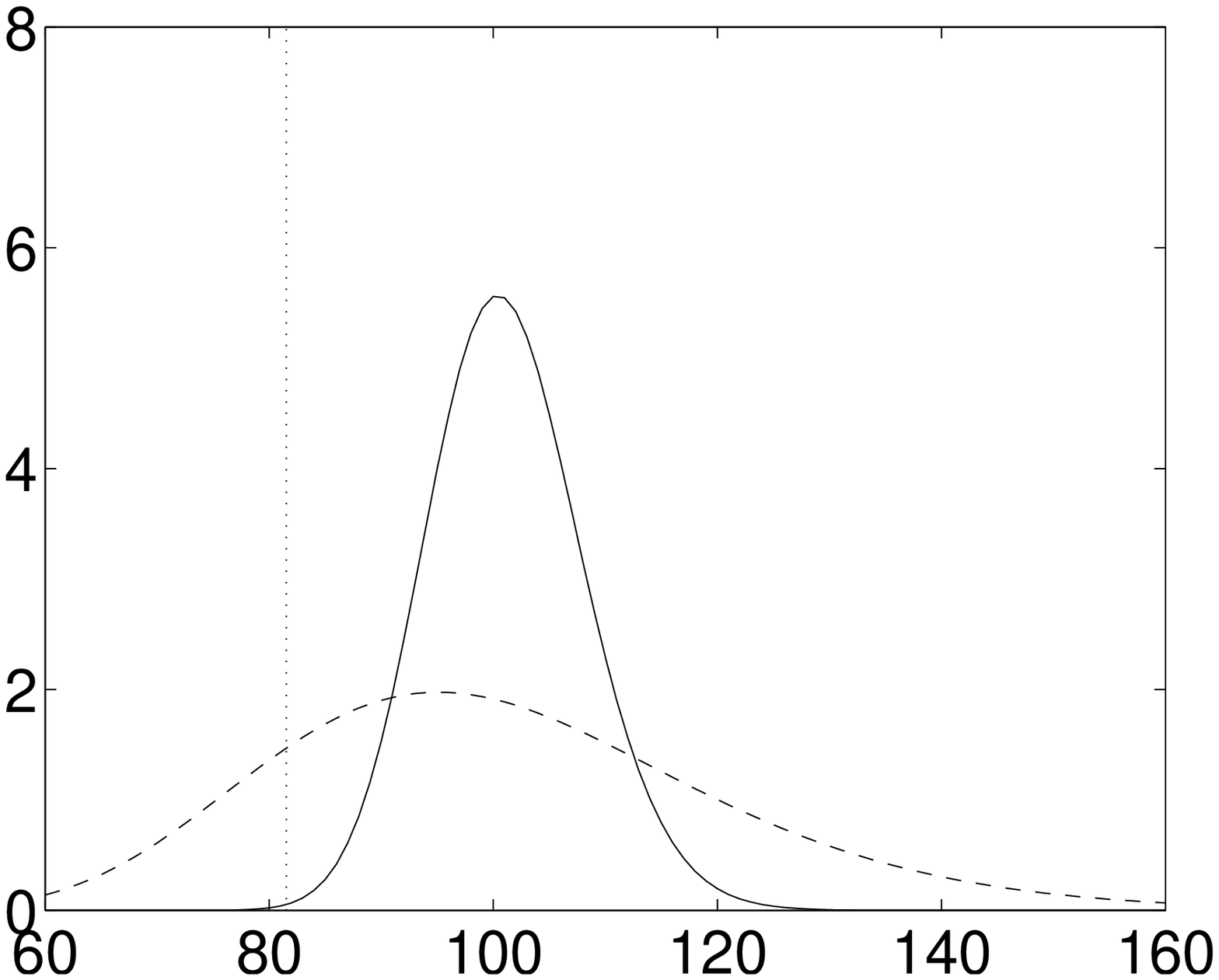}\\
\end{tabular}
  \caption{Non calibrated (dashed line) and calibrated (solid line) risk neutral distribution and price level (vertical dotted line) at last point of the sample, i.e. $t=504$, for different maturities (rows) and different volatility levels (columns).}\label{CalibrationSimDensity}
\end{figure}

The results of the calibration exercises are given in
Table \ref{TabMCMC} and Figure \ref{CalibrationSim}. Note the following salient features:
\begin{itemize}
\item The autoregressive coefficient is significant for all maturities. The proximity parameter is significant only for the last maturity. The value of the precision parameter increases with the maturity.
Figure \ref{CalibrationSim} shows the non-calibrated and calibrated PITs. Fig. \ref{CalibrationSimDensity} shows the predictive density, and the calibrated predictive, for the prices at time $t=504$ using the implied densities available at time $t-\tau_{j}$ for different $j$ (rows) and different wrong values of the volatility parameter $\sigma$ (columns).
\item We also consider a more parsimonious model, where we assume $\beta_{kj}=\beta_{k}$ and $\alpha_{kj}=\alpha_{k}$ for all $j=1,\ldots,M$. The results are given in Table \ref{TabMCMC}. 
\end{itemize}

\begin{table}[p]
\begin{tabular}{|c|cc|cc|cc|}
\multicolumn{7}{c}{Panel (a) ($\sigma=0.1$)}\\
\hline 
            &\multicolumn{2}{|c|}{$\tau_{j}$, $j=1$}&\multicolumn{2}{|c|}{$\tau_{j}$, $j=2$}&\multicolumn{2}{|c|}{$\tau_{j}$, $j=3$}\\
\hline
$\theta_{ij}$ &  $\hat{\theta}_{ij}$ & CI &   $\hat{\theta}_{ij}$ & CI &  $\hat{\theta}_{ij}$ & CI \\
\hline
$\gamma_{j}$&1.42&(1.01,1.51)&2.82&(2.79,2.93)&13.46&(13.40,13.64)\\
$\alpha_{0j}$&-0.32&(-0.44,-0.25)&-0.55&(-0.64,-0.46)&-1.09&(-1.15,-1.02)\\
$\alpha_{1j}$&0.43&(0.32,0.48)& 0.51&(0.35,0.61)&0.32&(0.23,0.42)\\
$\beta_{1j}$& & &0.11&(0.01,0.21)&0.16&(0.04,0.26)\\
$\beta_{2j}$&0.18&(0.06,0.27)&0.03&(0.01,0.15)& &\\
\hline
\multicolumn{7}{c}{\vspace{2pt}}\\
\multicolumn{7}{c}{Panel (b) ($\sigma=0.2$)}\\
\hline 
            &\multicolumn{2}{|c|}{$\tau_{j}$, $j=1$}&\multicolumn{2}{|c|}{$\tau_{j}$, $j=2$}&\multicolumn{2}{|c|}{$\tau_{j}$, $j=3$}\\
\hline
$\theta_{ij}$ &  $\hat{\theta}_{ij}$ & CI &   $\hat{\theta}_{ij}$ & CI &  $\hat{\theta}_{ij}$ & CI \\
\hline
$\gamma_j$   & 3.75&( 3.73, 3.81)& 7.01 &(6.67,7.23)   &14.03.88 &(13.83,14.16)\\
$\alpha_{0j}$&-0.24&(-0.34,-0.11)&-0.11 &(-0.19,-0.03) &-0.23    &(-0.29,-0.18)\\
$\alpha_{1j}$&0.37 &(0.23,0.47)  &0.30  &(0.24,0.41)   &0.47    &(0.32,0.58)  \\
$\beta_{1j}$ &     &             & 0.37&(0.27,0.43)    &0.05    &(-0.09,0.21)   \\
$\beta_{2j}$ &0.13 &(0.03,0.21)  &-0.02  &(-0.09,0.08)    &        &             \\
\hline
\end{tabular}
\begin{tabular}{|c|cc|}
\multicolumn{3}{c}{\vspace{2pt}}\\
\multicolumn{3}{c}{Panel (c)  ($\sigma=0.1$)}\\
\hline 
            &\multicolumn{2}{|c|}{$\tau_{j}, j=1,2,3$}\\
\hline
$\gamma$&99.4&(42.7,171.81)\\
$\alpha_{0}$&0.17&(-5.25,7.19)\\
$\alpha_{1}$&-0.02&(-8.37,5.21)\\
$\beta_{1}$&-0.74&(-6.79,5.98)\\
$\beta_{2}$&0.56&(-5.58,7.55)\\
\hline
\end{tabular}
\begin{tabular}{|c|cc|}
\multicolumn{3}{c}{\vspace{2pt}}\\
\multicolumn{3}{c}{Panel (d)  ($\sigma=0.2$)}\\
\hline 
            &\multicolumn{2}{|c|}{$\tau_{j}, j=1,2,3$}\\
\hline
$\gamma$    &95.3 &(49.57,159.66)\\
$\alpha_{0}$&1.55 &(-4.43,7.15)\\
$\alpha_{1}$&-0.48&(-5.8,4.17)\\
$\beta_{1}$ &-0.37&(-5.81,4.27)\\
$\beta_{2}$ &-0.04&(-7.48,7.77)\\
\hline
\end{tabular}
\caption{Posterior mean ($\hat{\theta}_{i}$) and 95\% credibility intervals (CI), for the parameters of the $\beta$-MRF. The non-calibrated predictive models with $\sigma=0.1$ (panels (a) for the hierarchical and (c) for the pooled model) and $\sigma=0.2$ (panels (b) for the hierarchical and (d) for the pooled model), when the true value of the scale parameter is $\sigma=0.15$.}\label{TabMCMC}
\end{table}

\section{Foreign exchange market application}\label{Section4}
We apply our methodology to over-the-counter annualized implied volatilities on the Eurodollar spot for different tenors (one month, two months, and six months), spanning from 01/01/2010
until 01/04/2013. For the computation of the risk neutral densities, we applied the same procedure consisting of first fitting a spline to the implied volatility for each tenor separately as in
\cite{PanigirtzoglouSkiadopoulos04,VergotePuigvert10}, in order to
transform back to the option price space and take the second
derivative to yield the risk neutral density\footnote{A more thorough description of how to
estimate the risk-neutral density obtained in our work is given in our appendix.}. For an extensive review
on how to extract risk-neutral densities from option prices with
Matlab code included, see \cite{FusaiRoncoroni00}. We apply this methodology both for the case where we assume
that each tenor has its own calibration function, and for the case
where we assume that there is a single calibration function that works
across several tenors by setting $\beta_{kj}=0$ in the specification of $\mu_{jt}$.

\begin{figure}[p]
\centering
\begin{tabular}{cc}
\multicolumn{2}{c}{One month}\\
  \includegraphics[width=170pt, height=150pt,
  angle=0]{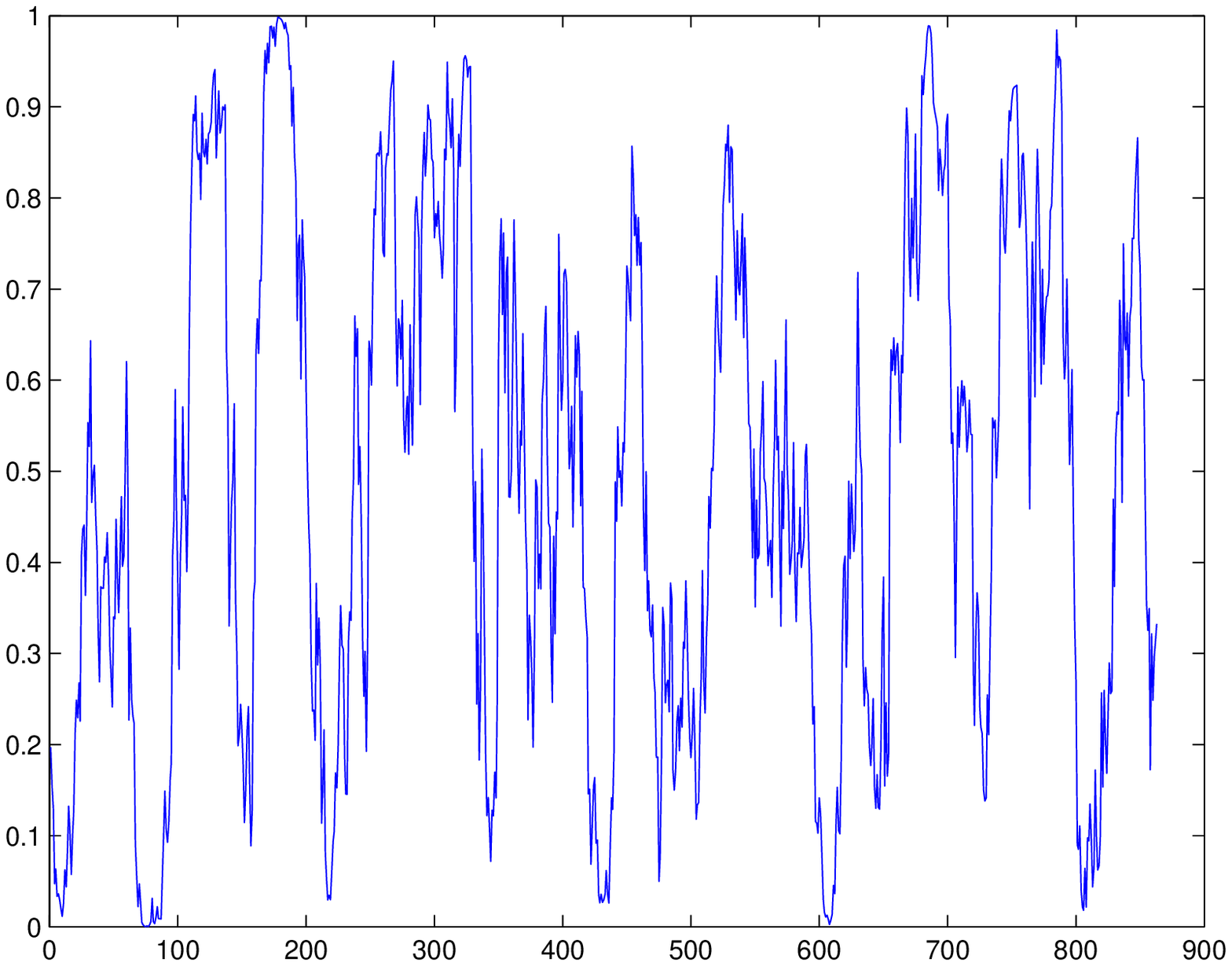}&  \includegraphics[width=170pt,
  height=150pt, angle=0]{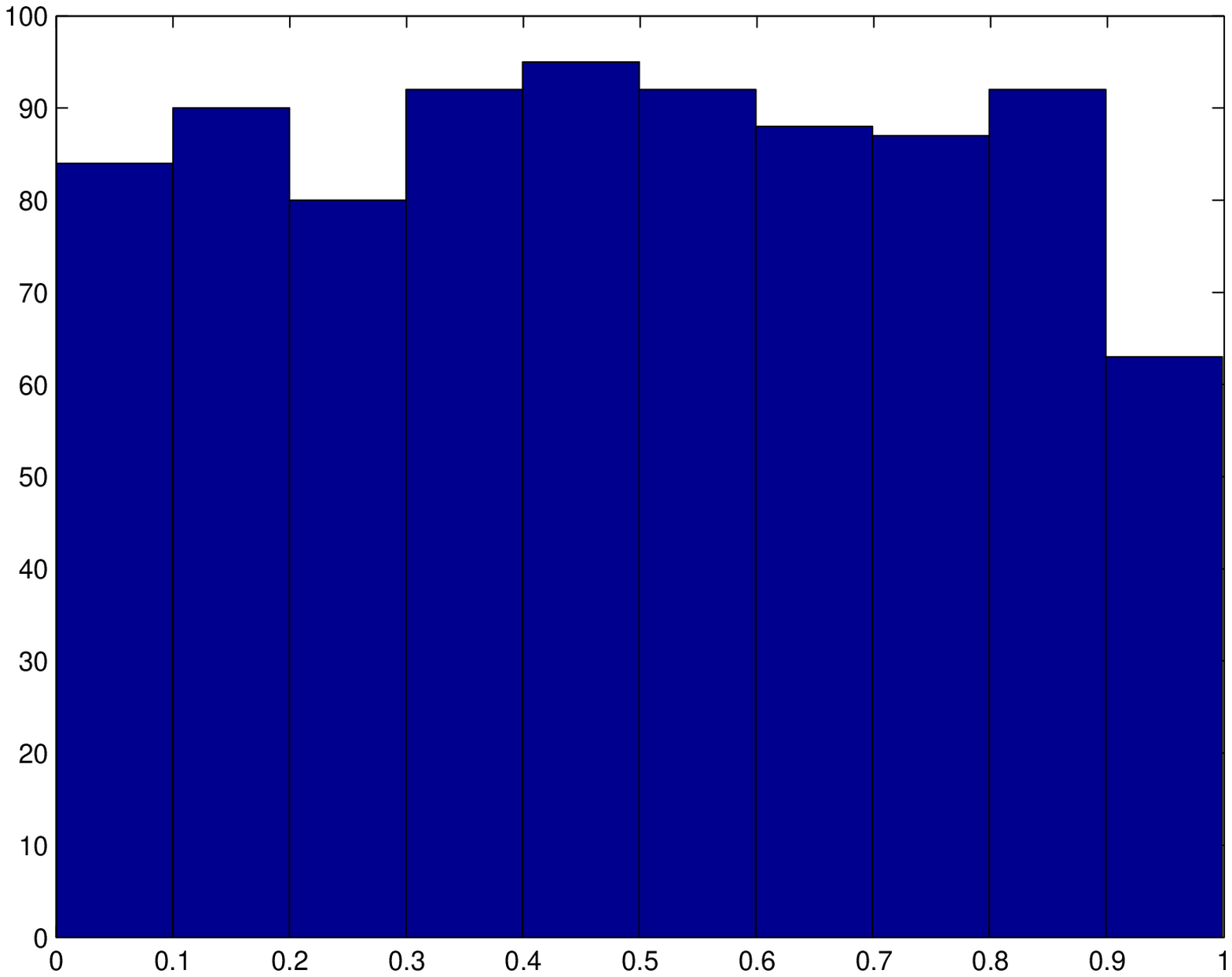}\\
\multicolumn{2}{c}{Two months}\\
  \includegraphics[width=170pt, height=150pt,
  angle=0]{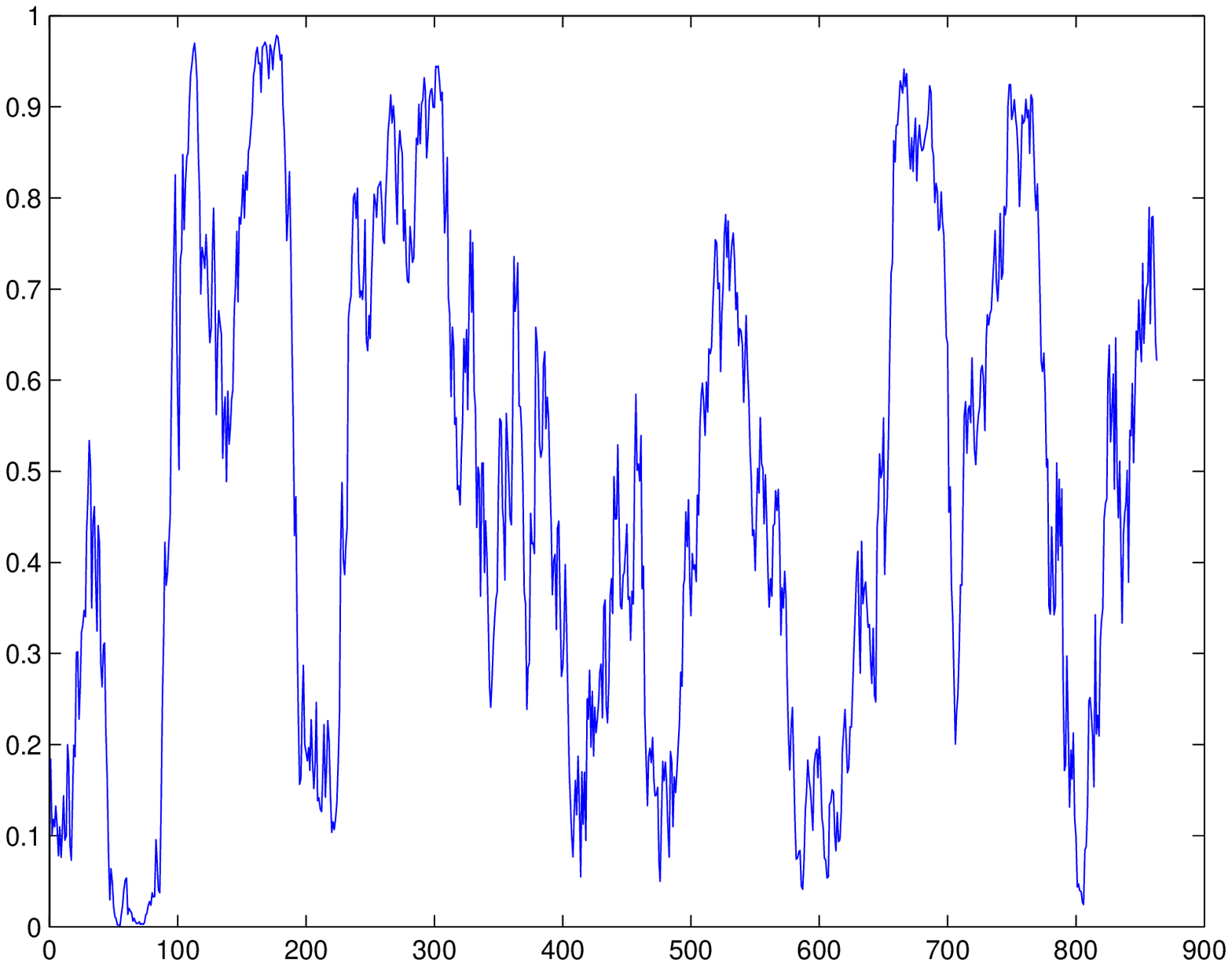}&  \includegraphics[width=170pt,
  height=150pt, angle=0]{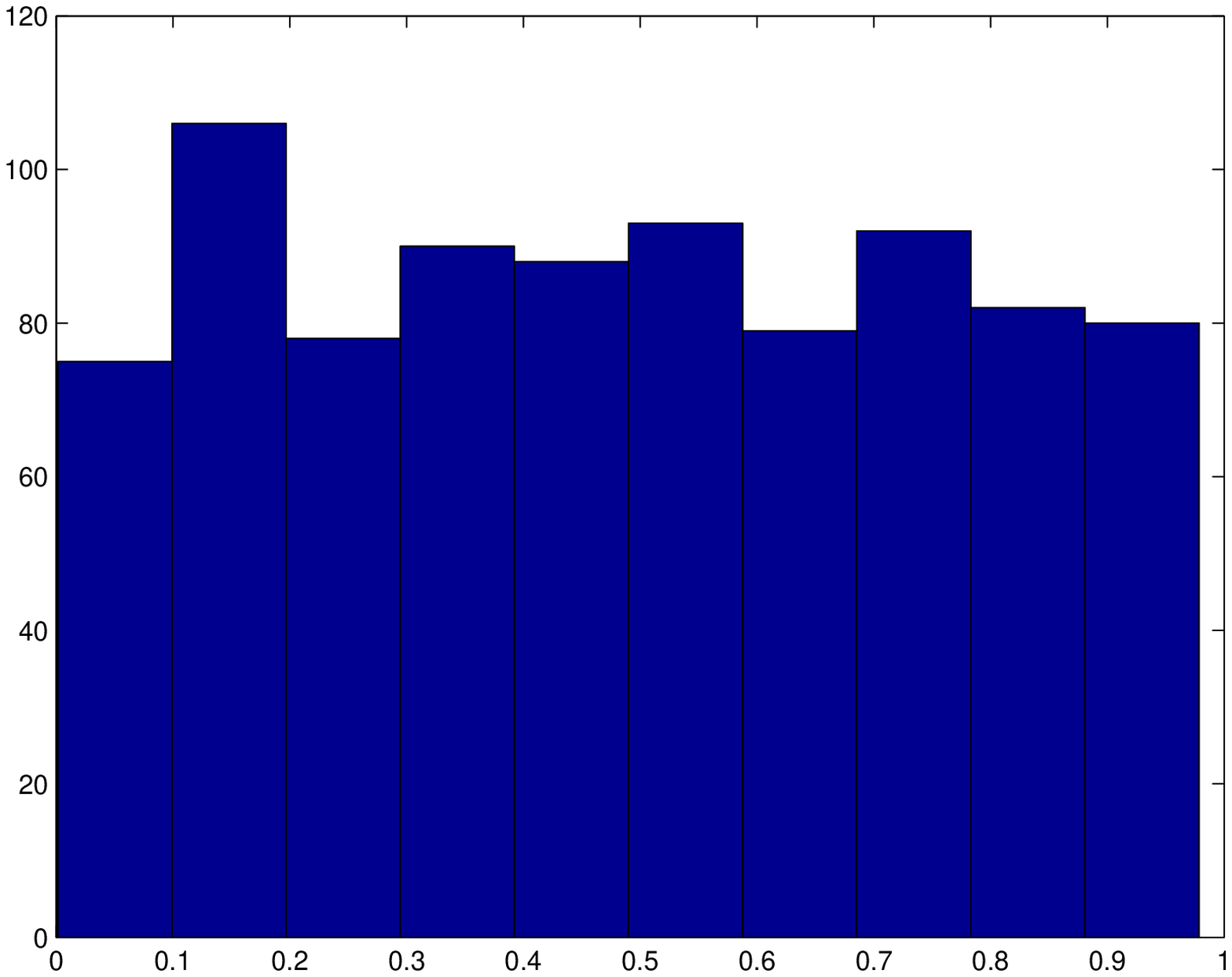}\\
\multicolumn{2}{c}{Six months}\\
  \includegraphics[width=170pt, height=150pt, angle=0]{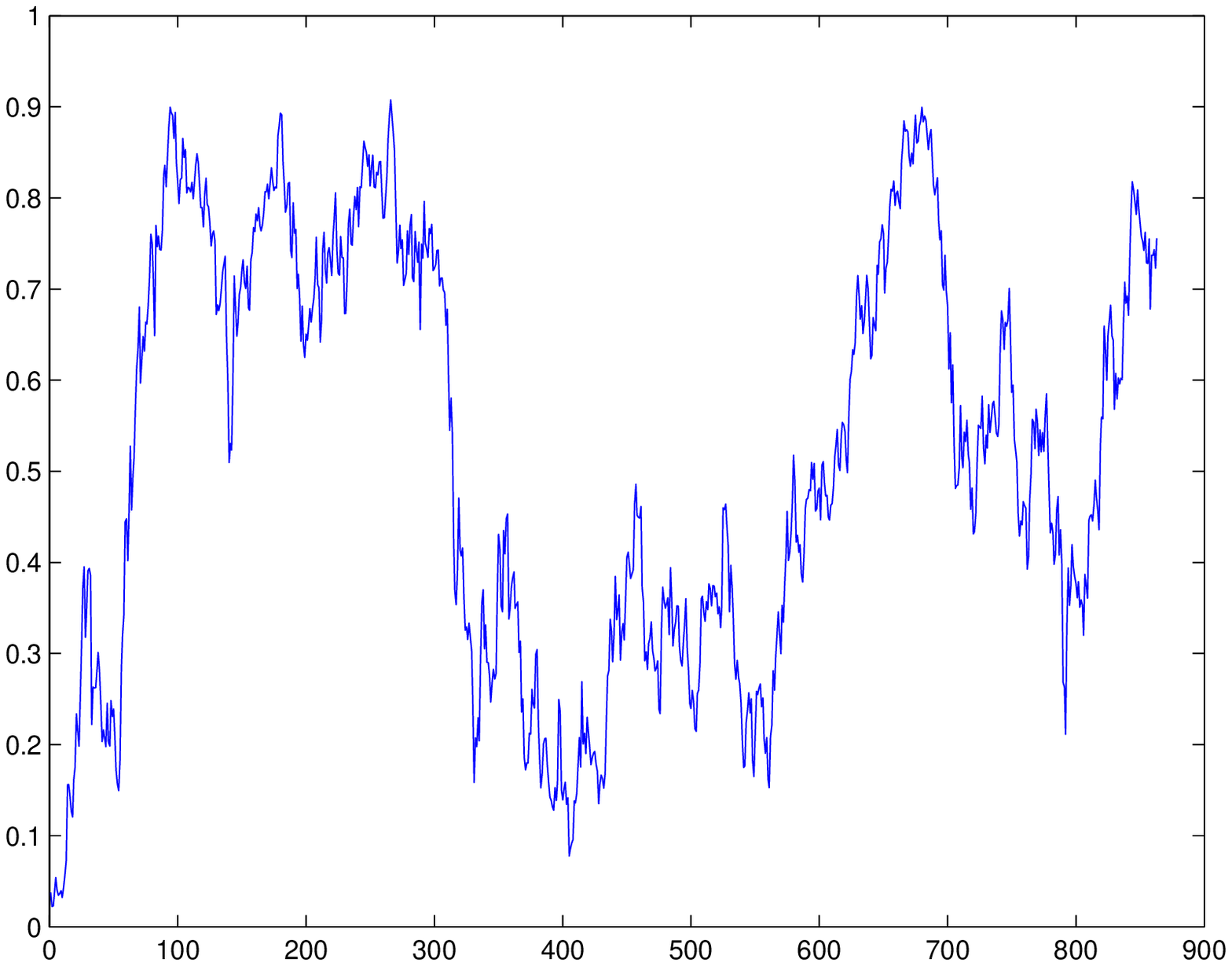}
&  \includegraphics[width=170pt, height=150pt, angle=0]{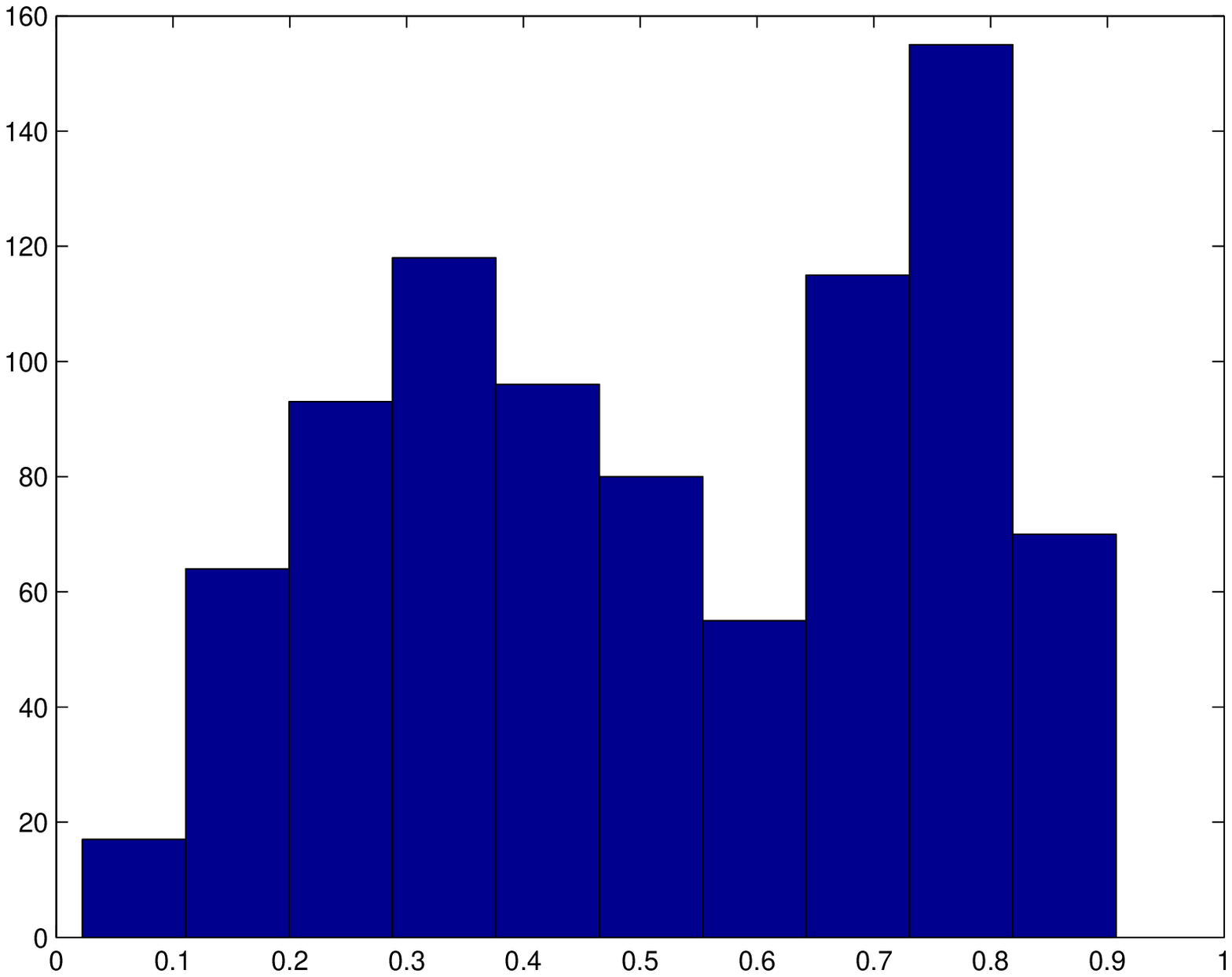}\\
\end{tabular}
  \caption{PIT time series (first column) and histogram (second column).}\label{PIT_Hist}
\end{figure}

Figure \ref{PIT_Hist} shows the time series (left column) and the histograms (right column) of the different PIT series. Even though most histograms of the time series of the PITs look very uniform due to the large sample size, the longer the tenor, the stronger the autoregressive component.


\begin{figure}[p]
\centering
\begin{tabular}{c}
  \includegraphics[width=170pt, height=150pt, angle=0]{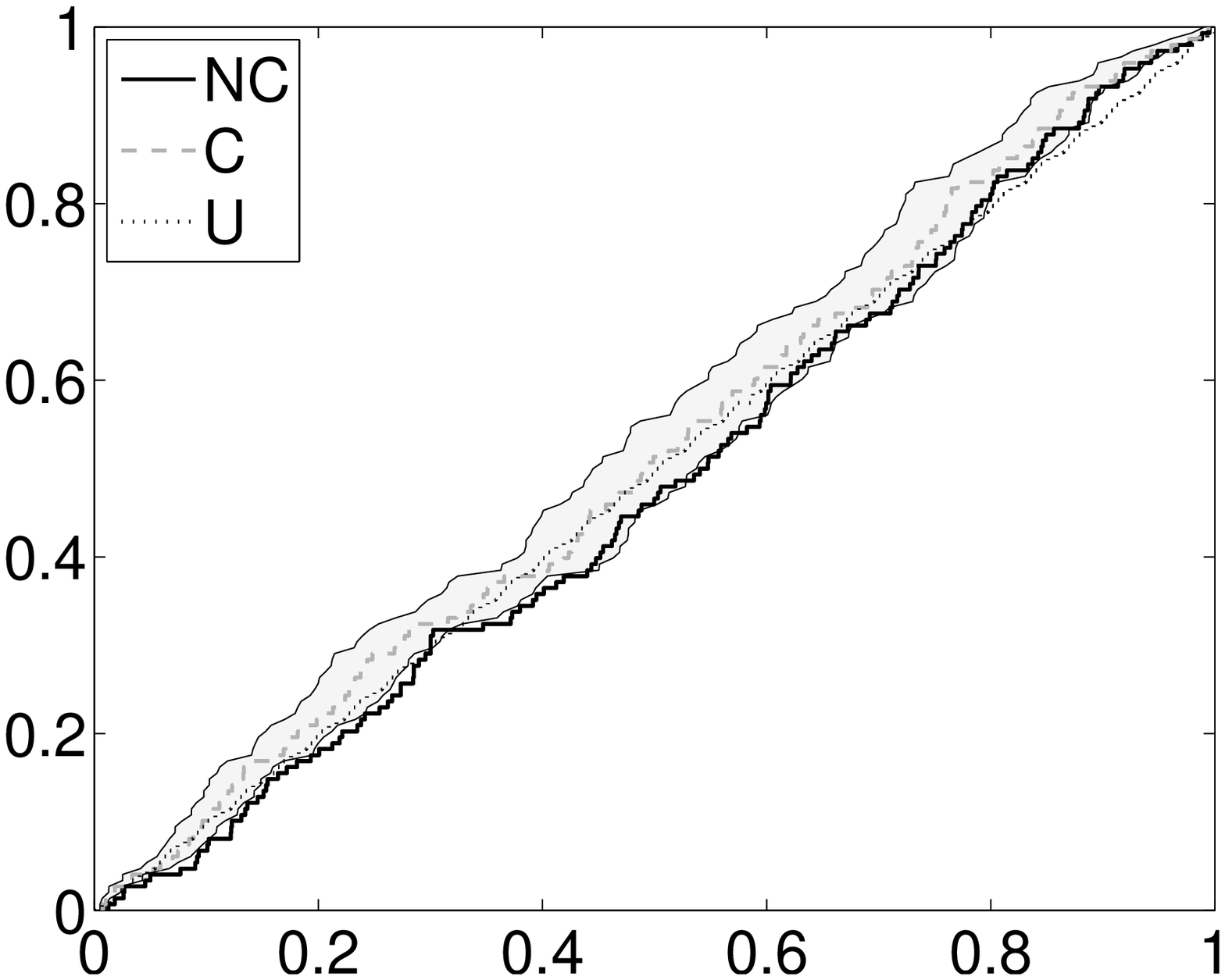}\\
  \includegraphics[width=170pt, height=150pt, angle=0]{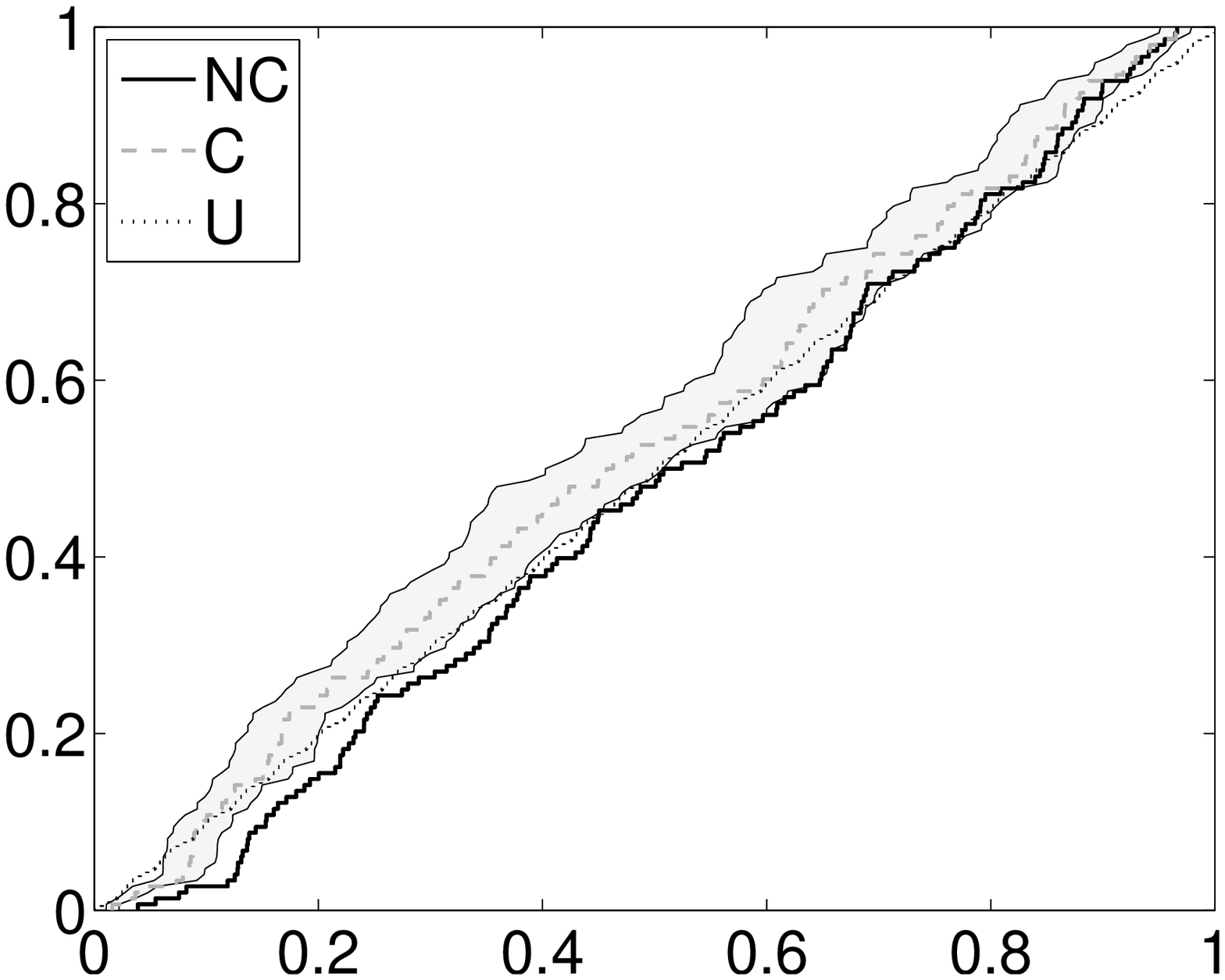}\\
  \includegraphics[width=170pt, height=150pt, angle=0]{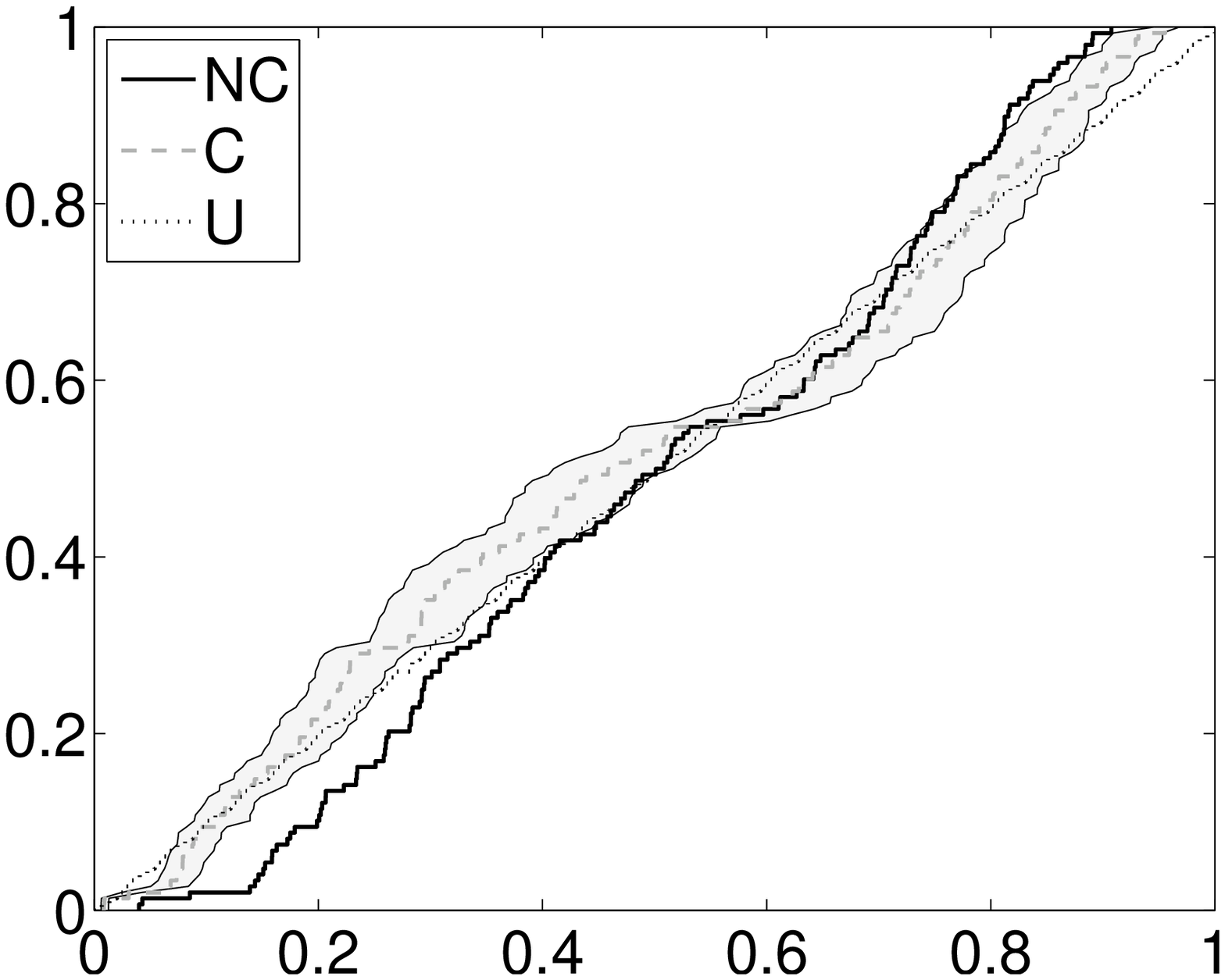}\\
\end{tabular}
  \caption{Non calibrated (solid line), perfectly calibrated (dotted line), and $\beta$-MRF calibrated (gray dashed line) risk neutral distributions for the three different maturities (rows): one, two and six months. In each plot, gray areas represent the 95\% HPD region.}\label{CalibrationData}
\end{figure}

We further display below the risk neutral densities estimated on the last day of the sample, 01/04/2013, for the different maturities, as well as their physical
densities, computed by applying the calibration function to each of the
risk-neutral densities. We apply our $\beta$-MRF calibration model with the prior and MCMC setting used in the simulated experiments (see previous section).
The results are given in Figure \ref{CalibrationData}. As it results from panel (a) in Table \ref{TabMCMCReal}, we found evidence of autocorrelation component (coefficient $\alpha_{1j}$) and of dependence across neighbouring maturities (coefficients $\beta_{ij}$). From panel (b) of the same figure one can see that the value of the autoregressive coefficient decreases when thinning (thinning factor 100/15) is applied to the PITs time series in order to reduce the dependence between the samples.

\begin{table}[p]
\begin{tabular}{|c|cc|cc|cc|}
\multicolumn{7}{c}{Panel (a) (Original data sample)}\\
\hline 
            &\multicolumn{2}{|c|}{$\tau_{j}$, $j=1$}&\multicolumn{2}{|c|}{$\tau_{j}$, $j=2$}&\multicolumn{2}{|c|}{$\tau_{j}$, $j=3$}\\
\hline
$\theta_{ij}$ &  $\hat{\theta}_{ij}$ & CI &   $\hat{\theta}_{ij}$ & CI &  $\hat{\theta}_{ij}$ & CI \\
\hline
$\gamma_{j}$ &  2.24& (2.14, 3.01)& 2.78&( 2.76, 2.98) & 3.55&( 3.42,3.97)\\
$\alpha_{0j}$& -0.04&(-0.16, 0.09)&-0.06&(-0.21, 0.07)&-0.08&(-0.24, 0.04)\\
$\alpha_{1j}$&  0.15& (0.06, 0.24)& 0.31&( 0.14, 0.45)& 0.31& (0.21, 0.52)\\
$\beta_{1j}$ &      &             & 0.14&( 0.04, 0.28)& 0.13& (-0.01, 0.26)\\
$\beta_{2j}$ &  0.17& (0.04, 0.28)& -0.01&(-0.2, 0.01)&     &             \\
\hline
\multicolumn{7}{c}{\vspace{2pt}}\\
\multicolumn{7}{c}{Panel (b) (Thinned data sample)}\\
\hline 
            &\multicolumn{2}{|c|}{$\tau_{j}$, $j=1$}&\multicolumn{2}{|c|}{$\tau_{j}$, $j=2$}&\multicolumn{2}{|c|}{$\tau_{j}$, $j=3$}\\
\hline
$\theta_{ij}$ &  $\hat{\theta}_{ij}$ & CI &   $\hat{\theta}_{ij}$ & CI &  $\hat{\theta}_{ij}$ & CI \\
\hline
$\gamma_j$   & 2.63 &( 2.52,2.78) &2.57&( 2.34,2.71)&3.52&( 3.48,3.62)\\
$\alpha_{0j}$& 0.03 &(-0.14,0.23) &0.04&(-0.15,0.18)&0.02&(-0.17,0.21)\\
$\alpha_{1j}$& 0.05 &(-0.25,0.21) &0.10&(-0.04,0.33)&0.11&(-0.02,0.23)\\
$\beta_{1j}$ &      &             &0.06&(0.01 ,0.26)&0.09&(-0.01,0.34)\\
$\beta_{2j}$ & 0.07 &(-0.10,0.32) &0.03&(-0.14,0.23)&    &            \\
\hline
\end{tabular}
\caption{Posterior mean ($\hat{\theta}_{i}$) and 95\% credibility intervals (CI), for the parameters of the $\beta$-MRF. The non-calibrated and $\beta$-MRF calibrated predictive pits empirical distribution function for original data (panel (a)) and thinned data (panel (b)), with thinning factor 100/15.}\label{TabMCMCReal}
\end{table}

\section{Conclusion}\label{Section5}
This paper, which builds on the methodology by \cite{CasDalLei12}, provides a new modelling framework using both the derivative implied volatilities (at-the-money, mid-market, end-of-day volatilities)
and synchronized spot and forwards for the term structure of the implied probability,
which accounts for the possible dependence between PITs at different
maturities, and different dates for a given maturity. This approach allows borrowing of information between the different tenors, for both the risk-neutral and the physical
measures. We also provide a proper inferential Bayesian framework that
allows the inclusion of parameter uncertainty in the density calibration functions, normally a factor overseen in the literature, and therefore also in the physical densities. 

Modelling
the time evolution of the predictive densities and the relationship
between densities from many sources is a challenging issue.  For
example, in traditional approaches, when reconstructing the calibration function there cannot be
any overlapping time intervals so that the PITs are independent in
order to estimate the beta calibration function as explained in
\cite{FacKin90} and later used in
\citep{VergotePuigvert10,VeselaPuigvert13}. Using independent PITs has
the drawback of requiring large amounts of data, not always available for new assets or assets that do not trade frequently,  to have a reliable calibration
function. Our methodology takes advantage from using all information available, without thinning of the source, in a
rolling window fashion, since the induced correlation is incorporated in the modelling of the dependent PITs.

Future research will include adapting our methodology to other
assets, such as stocks, commodities, fixed income indices, and other exchange-traded markets, which trade on fixed expiry contracts rather than rolling constant term contracts, 
by interpolating the risk neutral densities for different constant maturities from fixed expiry contracts as
done in \citep{VergotePuigvert10}.

\section*{Acknowledgments}
\noindent We thank seminar participants at the George Box Workshop
2014, Mathematics Department of Universidad Nacional of Medellin,
Colombia, 2014. Roberto Casarin's research is supported by funding from the European Union, Seventh Framework Programme FP7/2007-2013 under grant agreement SYRTO-SSH-2012-320270, and by the Italian Ministry of Education, University and Research (MIUR) PRIN 2010-11 grant MISURA.


\newpage
\appendix\section{Risk Neutral density estimation}
Our application consists of the daily implied annualized volatility on
the Eurodollar for different (constant maturity, rather than fixed day of expiry) expirations. The full data consists of the closing snapshots at the end of the
London business day for spot, forwards and at-the-money implied volatilities for the period of the 01/01/2010 to the
01/04/2013. We follow \cite{CampaChangReider97,VergotePuigvert10} and
transform the option prices (y-axis) and strikes (x-axis) to the sigma
(y-axis) and delta (x-axis) space in order to fit a cubic smoothing
spline to the volatility smile. The reason for working in the
sigma-delta space instead of the regular option price space is that undesired noise in
the option data is introduced through high liquidity and transaction
volumes which then makes difficult the interpolation of option
prices. By fitting the implied volatility (sigma-delta) instead of the option prices
directly, one is able to circumvent the latter problem of the noise in
the option data by \cite{Shimko93,HutchinsonLoPoggio94,Malz97,SahaliaLo98,EngleRosenberg00,BlissPanigirtzoglou02}. Additionally, this is the natural approach in over-the-counter options markets, where option prices are often quoted by market-makers in volatility space, rather than price space.

Using the same notation as in \cite{VergotePuigvert10}, the optimal cubic smoothing spline of the implied
volatility is the one that minimizes the following function:
\begin{equation}
\min \lambda\sum_{i=1}^{n}\omega_{i}(\sigma_{i}-\hat{\sigma(\Theta)_{i}})^{2}+(1-\lambda)\int_{0}^{1}g^{''}(\delta,\Theta)d\delta
\end{equation}
where $\delta$ is the partial derivative of the Black and Scholes option
call price with respect to the underlying, $\sigma_{i}$,
$\hat{\sigma(\Theta)}$, and $\omega_{i}=\frac{\nu_{i}}{\sum_{i}^{n}\nu_{i}}$ are the observed volatility,
fitted volatility, and weight of observation $i$, together with its
Greek Black and Scholes vega $\nu$.  
respectively. Furthermore, $\Theta$ represents the matrix of polynomial
parameters of the cubic spline while $g()$ is the cubic spline
function. The value for $\lambda$ used is equal to 0.99. It is
worthwhile noting that the Black-Scholes formula
is used solely to convert the option prices into/from their
implied volatilities, in order to make the smoothing more
effectively. This approach does not imply that we are assuming the Black and Scholes pricing formula is the correct one, but is only a way to make the smoothing more effective in the
interpolation\footnote{The function csaps was used to perform cubic smoothing
  spline interpolation in Matlab}.
%
\newpage
\bibliographystyle{ims}
\bibliography{References}

\begin{thebibliography}{45}
\expandafter\ifx\csname natexlab\endcsname\relax\def\natexlab#1{#1}\fi
\expandafter\ifx\csname url\endcsname\relax
  \def\url#1{\texttt{#1}}\fi
\expandafter\ifx\csname urlprefix\endcsname\relax\def\urlprefix{URL }\fi
\providecommand{\eprint}[2][]{\url{#2}}

\bibitem[{A\"{i}t-Sahalia and Duarte(2003)}]{Ait03}
\textsc{A\"{i}t-Sahalia, Y.} and \textsc{Duarte, J.} (2003).
\newblock Nonparametric option pricing under shape restrictions.
\newblock \textit{Journal of Econometrics}, \textbf{116} 9--47.

\bibitem[{Ait-Sahalia and Lo(1998)}]{SahaliaLo98}
\textsc{Ait-Sahalia, Y.} and \textsc{Lo, A.} (1998).
\newblock Nonparametric estimation of state-price densities implicit in
  financial asset prices.
\newblock \textit{Journal of Finance}, \textbf{53(2)} 449--547.

\bibitem[{Bhar and Chiarella(2000)}]{BhaChi00}
\textsc{Bhar, R.} and \textsc{Chiarella, C.} (2000).
\newblock Expectations of monetary policy in {A}ustralia implied by the
  probability distribution of interest rate derivatives.
\newblock \textit{European Journal of Finance}, \textbf{6} 113--125.

\bibitem[{Billio et~al.(2013)Billio, Casarin, Ravazzolo and
  Van~Dijk}]{BilCasRavVan13}
\textsc{Billio, M.}, \textsc{Casarin, R.}, \textsc{Ravazzolo, F.} and
  \textsc{Van~Dijk, H.} (2013).
\newblock Time-varying combinations of predictive densities using nonlinear
  filtering.
\newblock \textit{Journal of Econometrics}, \textbf{177} 213--232.

\bibitem[{Black and Scholes(1973)}]{bs:1973}
\textsc{Black, F.} and \textsc{Scholes, M.} (1973).
\newblock {T}he pricing of options and corporate liabilities.
\newblock \textit{Journal of Political Economy}, \textbf{81} 637--654.

\bibitem[{Bliss and Panigirtzoglou(2002)}]{BlissPanigirtzoglou02}
\textsc{Bliss, R.} and \textsc{Panigirtzoglou, N.} (2002).
\newblock Testing the stability of implied probability density functions.
\newblock \textit{Journal of Banking and Finance}, \textbf{26(2)} 381--422.

\bibitem[{Bliss and Panigirtzoglou(2004)}]{BlissPanigirtzoglou04}
\textsc{Bliss, R.} and \textsc{Panigirtzoglou, N.} (2004).
\newblock Option-implied risk aversion estimates.
\newblock \textit{Journal of Finance}, \textbf{59} 407--446.

\bibitem[{Bouguila et~al.(2006)Bouguila, Ziou and Monga}]{BouZioMon06}
\textsc{Bouguila, N.}, \textsc{Ziou, D.} and \textsc{Monga, E.} (2006).
\newblock Practical {B}ayesian estimation of a finite beta mixture through
  {G}ibbs sampling and its applications.
\newblock \textit{Statistics and Computing}, \textbf{16} 215--225.

\bibitem[{Boyarchenko and Levendorskii(2002)}]{BoyarchenkoLevendorskii02}
\textsc{Boyarchenko, S.~I.} and \textsc{Levendorskii, S.~Z.} (2002).
\newblock \textit{Non-Gaussian Merton-Black-Scholes theory}.
\newblock Wold Scientific. Advanced Series on Statistical Science and Applied
  Probability-Vol 9.

\bibitem[{Branscum et~al.(2007)Branscum, Johnson and Thurmond}]{BraJohThu07}
\textsc{Branscum, A.~J.}, \textsc{Johnson, W.} and \textsc{Thurmond, M.}
  (2007).
\newblock Bayesian beta regresssion: {A}pplications to household expenditure
  data and genetic distance between foot-and-mouth disease viruses.
\newblock \textit{Australian \& New Zealand Journal of Statistics}, \textbf{49}
  287--301.

\bibitem[{Bremaud(1999)}]{Bre99}
\textsc{Bremaud, P.} (1999).
\newblock \textit{Markov {C}hains: {G}ibbs Fields, Monte Carlo Simulation, and
  Queues}.
\newblock Springer.

\bibitem[{Campa et~al.(1997)Campa, Chang and Reider}]{CampaChangReider97}
\textsc{Campa, J.}, \textsc{Chang, P.} and \textsc{Reider, R.} (1997).
\newblock {ERM} bandwidths for {EMU} and after: evidence from foreign exchange
  options.
\newblock \textit{Economic Policy}, \textbf{24} 55--89.

\bibitem[{Carlson et~al.(2005)Carlson, Craig and Melick}]{CarCraMel05}
\textsc{Carlson, J.~B.}, \textsc{Craig, B.} and \textsc{Melick, W.~R.} (2005).
\newblock Recovering market expectations of {FOMC} rate changes with options on
  federal funds futures.
\newblock \textit{Journal of Futures Markets}, \textbf{25} 1203--1242.

\bibitem[{Casarin et~al.(2012)Casarin, Valle and Leisen}]{CasDalLei12}
\textsc{Casarin, R.}, \textsc{Valle, L.~D.} and \textsc{Leisen, F.} (2012).
\newblock Bayesian model selection for beta autoregressive processes.
\newblock \textit{Bayesian Analysis}, \textbf{7} 1--26.

\bibitem[{Delbaen and Schachermayer(2011)}]{DelbaenSchachermayer11}
\textsc{Delbaen, F.} and \textsc{Schachermayer, W.} (2011).
\newblock \textit{The Mathematics of Arbitrage}.
\newblock Springer Finance.

\bibitem[{Engle and Rosenberg(2000)}]{EngleRosenberg00}
\textsc{Engle, R.} and \textsc{Rosenberg, J.} (2000).
\newblock Testing the volatility term structure using option hedging criteria.
\newblock \textit{Journal of Derivatives}, \textbf{8} 10--29.

\bibitem[{Esscher(1932)}]{Esscher32}
\textsc{Esscher, F.} (1932).
\newblock On the probability function in the collective theory of risk.
\newblock \textit{Skandinavisk Aktuarietidskrift}, \textbf{15} 175--195.

\bibitem[{Fackler and King(1990)}]{FacKin90}
\textsc{Fackler, P.~L.} and \textsc{King, R.~P.} (1990).
\newblock Calibration of option-based probability assesments in agricoltural
  commodity markets.
\newblock \textit{American Journal of Agricultural Economics}, \textbf{72}
  73--83.

\bibitem[{Fawcett et~al.(2013)Fawcett, Kapetanios, Mitchell and
  Price}]{MitKap13}
\textsc{Fawcett, N.}, \textsc{Kapetanios, G.}, \textsc{Mitchell, J.} and
  \textsc{Price, S.} (2013).
\newblock Generalised density forecast combinations.
\newblock Tech. rep., Warwick Business School.

\bibitem[{Fusai and Roncoroni(2000)}]{FusaiRoncoroni00}
\textsc{Fusai, G.} and \textsc{Roncoroni, A.} (2000).
\newblock \textit{Implementing Models in Quantitative Finance: Methods and
  Cases}.
\newblock Springer.

\bibitem[{Gerber and Shiu(1994)}]{GerberShiu94}
\textsc{Gerber, H.~U.} and \textsc{Shiu, S.} (1994).
\newblock Option pricing by {E}sscher transforms.
\newblock \textit{Transactions of the Society of Actuaries}, \textbf{46}
  99--191.

\bibitem[{Geweke(1992)}]{GJ92}
\textsc{Geweke, J.} (1992).
\newblock Evaluating the accuracy of sampling-based approaches to the
  calculation of posterior moments.
\newblock In \textit{In Bayesian Statistics}. University Press, 169--193.

\bibitem[{Geweke and Amisano(2011)}]{GewekeAmisano2011}
\textsc{Geweke, J.} and \textsc{Amisano, G.} (2011).
\newblock Optimal prediction pools.
\newblock \textit{Journal of Econometrics}, \textbf{164} 130--141.

\bibitem[{Gneiting et~al.(2005)Gneiting, Raftery, Westveld and
  Goldman}]{GneRafWesGol05}
\textsc{Gneiting, T.}, \textsc{Raftery, A.}, \textsc{Westveld, A.~H.} and
  \textsc{Goldman, T.} (2005).
\newblock Calibrated probabilistic forecasting using ensemble model output
  statistics and minimum {CRPS} estimation.
\newblock \textit{Monthly Weather Review}, \textbf{133} 1098--1118.

\bibitem[{Gneiting and Ranjan(2013)}]{GneRan13}
\textsc{Gneiting, T.} and \textsc{Ranjan, R.} (2013).
\newblock Combining predictive distributions.
\newblock \textit{Electronic Journal of Statistics}, \textbf{7} 1747--1782.

\bibitem[{Hall and Mitchell(2007)}]{HallMitchell2007}
\textsc{Hall, S.~G.} and \textsc{Mitchell, J.} (2007).
\newblock Combining density forecasts.
\newblock \textit{International Journal of Forecasting}, \textbf{23} 1--13.

\bibitem[{Hutchinson et~al.(1994)Hutchinson, Lo and
  Poggio}]{HutchinsonLoPoggio94}
\textsc{Hutchinson, J.}, \textsc{Lo, A.} and \textsc{Poggio, T.} (1994).
\newblock A nonparametric approach to pricing and hedging derivative securities
  via learning networks.
\newblock \textit{Journal of Finance}, \textbf{49(3)} 851--889.

\bibitem[{Lai(2011)}]{WanNiLai11}
\textsc{Lai, W.-N.} (2011).
\newblock Comparison of methods to estimate option implied risk-neutral
  densities.
\newblock \textit{Quantitative Finance} 1--17.

\bibitem[{Liang(2010)}]{Lia10}
\textsc{Liang, F.} (2010).
\newblock A double {M}etropolis {H}astings sampler for spatial models with
  intractable normalizing constants.
\newblock \textit{Journal of Statistical Computation and Simulation},
  \textbf{80(9)} 1007--1022.

\bibitem[{Malz(1997)}]{Malz97}
\textsc{Malz, A.} (1997).
\newblock Estimating the probability distribution of the future exchange rate
  from option prices.
\newblock \textit{Journal of Derivatives}, \textbf{5(2)} 18--36.

\bibitem[{Merton(1973)}]{me:1973}
\textsc{Merton, R.} (1973).
\newblock {T}heory of rational option pricing.
\newblock \textit{Bell Journal of Economics and Management Science}, \textbf{4}
  141--183.

\bibitem[{M{\o}ller et~al.(2006)M{\o}ller, Pettitt, Reeves and
  Berthelsen}]{MolPetReeBer06}
\textsc{M{\o}ller, J.}, \textsc{Pettitt, A.~N.}, \textsc{Reeves, R.} and
  \textsc{Berthelsen, K.~K.} (2006).
\newblock An efficient {M}arkov chain {M}onte {C}arlo method for distributions
  with intractable normalizing constants.
\newblock \textit{Biometrika}, \textbf{93} 451--458.

\bibitem[{Murray et~al.(2006)Murray, Ghahramani and MacKay}]{MurGhaMac06}
\textsc{Murray, I.}, \textsc{Ghahramani, Z.} and \textsc{MacKay, D. J.~C.}
  (2006).
\newblock {MCMC} for doubly intractable distributions.
\newblock In \textit{Proceedings of the 22nd annual conference on uncertainty
  in Artificial Intelligence}.

\bibitem[{Nikodym(1930)}]{Radon}
\textsc{Nikodym, O.} (1930).
\newblock Sur une g\'{e}n\'{e}ralisation des int\'{e}grales de m. j. radon.
\newblock \textit{Fundamenta Mathematicae (in French) JFM 56.0922.02. Retrieved
  2009-05-11}, \textbf{15} 131--179.

\bibitem[{Panigirtzoglou and Skiadopoulos(2004)}]{PanigirtzoglouSkiadopoulos04}
\textsc{Panigirtzoglou, N.} and \textsc{Skiadopoulos, G.} (2004).
\newblock A new approach to modeling the dynamics of implied distributions:
  Theory and evidence from the {SP500} options.
\newblock \textit{Journal of Banking and Finance}.

\bibitem[{Robert and Rousseau(2002)}]{RobRou02}
\textsc{Robert, C.~P.} and \textsc{Rousseau, J.} (2002).
\newblock A mixture approach to {B}ayesian goodness of fit.
\newblock Technical Report 02009, CEREMADE, University Paris Dauphine.

\bibitem[{Rodriguez and ter Horst(2008)}]{RodriguezterHorst08}
\textsc{Rodriguez, A.} and \textsc{ter Horst, E.} (2008).
\newblock Bayesian dynamic density estimation.
\newblock \textit{Bayesian Analysis}, \textbf{3} 339--366.

\bibitem[{Rosenthal(2011)}]{Ros11}
\textsc{Rosenthal, J.~S.} (2011).
\newblock Optimal {P}roposal {D}istributions and {A}daptive {MCMC}.
\newblock In \textit{Handbook of {M}arkov {C}hain {M}onte {C}arlo: {M}ethods
  and {A}pplications} (G.~J. S.P.~Brooks, A.~Gelman and X.-L. Meng, eds.).
  Chapman \& Hall, 93--112.

\bibitem[{Rouah and Vainberg(2007)}]{RouahVainberg07}
\textsc{Rouah, F.} and \textsc{Vainberg, G.} (2007).
\newblock Option pricing models and volatility using excel-vba.
\newblock \textit{John Wiley and Sons}.

\bibitem[{Shimko(1993)}]{Shimko93}
\textsc{Shimko, D.} (1993).
\newblock Bounds of probability.
\newblock \textit{Risk}, \textbf{6(4)} 33--37.

\bibitem[{Sihvonen and V\"ah\"amaa(2014)}]{sihvah14}
\textsc{Sihvonen, J.} and \textsc{V\"ah\"amaa, S.} (2014).
\newblock Forward-looking monetary policy rules and option implied interest
  rate expectations.
\newblock \textit{The Journal of Futures Markets}, \textbf{34(4)} 346--373.

\bibitem[{Tankov and Cont(2003)}]{TankovCont03}
\textsc{Tankov, P.} and \textsc{Cont, R.} (2003).
\newblock \textit{Financial Modelling with Jump Processes}.
\newblock Chapman \& Hall/CRC.

\bibitem[{Vergote and Guti\'errez(2012)}]{VergotePuigvert10}
\textsc{Vergote, O.} and \textsc{Guti\'errez, J. M.~P.} (2012).
\newblock Interest rate expectations and uncertainty during {ECB} governing
  council days: evidence from intraday implied densities of 3-month {E}uribor.
\newblock \textit{Journal of Banking \& Finance}, \textbf{36} 2804--2823.

\bibitem[{Vesela and Guti\'errez(2013)}]{VeselaPuigvert13}
\textsc{Vesela, I.} and \textsc{Guti\'errez, J.~P.} (2013).
\newblock Getting real forecasts, state price densities and risk premium from
  {E}uribor options.
\newblock \textit{Available at SSRN: http://ssrn.com/abstract=2178428}.

\bibitem[{Vincent-Humphreys and Noss(2012)}]{devnos12}
\textsc{Vincent-Humphreys, R.~D.} and \textsc{Noss, J.} (2012).
\newblock Estimating probability distributions of future asset prices:
  empirical transformations from option-implied risk-neutral to real-world
  density functions.
\newblock Working Paper No. 455, Bank of England.

\end{thebibliography}

\end{document}